\documentclass[fleqn,10pt]{wlscirep}
\usepackage[utf8]{inputenc}

\usepackage{cite}
\usepackage{amsmath,amssymb,amsfonts}
\usepackage{algorithmic}
\usepackage{textcomp}
\usepackage{mathtools}
\usepackage{siunitx}
\usepackage{array}
\usepackage{multirow}
\usepackage{bm}
\usepackage{booktabs}
\usepackage[nolist]{acronym}
\usepackage{hyperref}
\usepackage{longtable}
\usepackage{booktabs}
\usepackage{multirow}
\usepackage{siunitx}
\usepackage{array}
\usepackage{makecell}
\usepackage{threeparttable}
\usepackage{subcaption}
\usepackage{enumitem}
\usepackage{tikz}
\usepackage{placeins}
\usepackage{xurl}

\setlist[itemize]{itemsep=1pt, topsep=1pt, parsep=1pt}

\makeatletter

\immediate\write18{python ./figures/ppt_to_pdf.py}

\DeclareSIUnit{\pu}{pu}
\DeclareSIUnit{\VA}{VA}
\DeclareSIUnit{\VAr}{VAr}
\newcommand\setcurrentname[1]{\def\@currentlabelname{#1}}

\title{A simulation based dataset of faults and events for machine learning in power systems}

\author[1,*]{Georg Kordowich}
\author[1]{Jonathan Loebel}
\author[2]{Julian Oelhaf}
\author[2]{Andreas Maier}
\author[2]{Siming Bayer}
\author[3]{Christian Bergler}
\author[1]{Johann Jaeger}

\affil[1]{Institute of Electrical Energy Systems, Friedrich-Alexander-Universität Erlangen-Nürnberg, Erlangen, Germany}
\affil[2]{Pattern Recognition Lab, Department of Computer Science, Friedrich-Alexander-Universität Erlangen-Nürnberg, Erlangen, Germany}
\affil[3]{Department of Electrical Engineering, Media and Computer Science, Ostbayerische Technische Hochschule Amberg-Weiden, Amberg, Germany}
\affil[*]{georg.kordowich@fau.de}

\begin{abstract}
The integration of inverter-based renewable energy sources into electric grids challenges conventional power system protection. Machine learning-based solutions can address these challenges by utilizing available data in modern smart grids. However, the lack of open datasets prevents reproducibility and fair comparisons between different approaches and their results, which hinders further progress. Therefore, this paper presents EvEMTBench, a synthetic dataset of faults and events generated using electromagnetic transient simulations. The physical plausibility of the power system simulation is ensured by validating the simulation parameters against established literature and providing comprehensive documentation of the simulation procedure. The dataset is designed for training, fine-tuning, and benchmarking machine learning models by providing synchronized point-on-wave voltage and current measurements at 9600 Hz across a diverse set of topologies and voltage levels. The inclusion of a wide range of fault and operating events allows the utilization of EvEMTBench for different tasks like incipient fault detection, fault localization, or event detection.
\end{abstract}

\begin{document}
\def\figureautorefname{Fig.}
\def\equationautorefname{Eq.}
\def\tableautorefname{Tab.}

\providecommand*\@fourthoffive[5]{#4}
\def\@getautoreftype#1.#2\@nil{#1}

\let\orig@autoref\autoref
\renewcommand*{\autoref}[1]{%
	\expandafter\ifx\csname r@#1\endcsname\relax
	\orig@autoref{#1}%
	\else
	\expandafter\let\expandafter\@ref@macro\csname r@#1\endcsname
	\edef\@ref@anchor{\expandafter\@fourthoffive\@ref@macro}%
	\edef\@ref@type{\expandafter\@getautoreftype\@ref@anchor.\@nil}%
	\expandafter\ifx\csname\@ref@type autorefname\endcsname\relax
	\orig@autoref{#1}%
	\else
	\csname\@ref@type autorefname\endcsname\nobreakspace\ref{#1}%
	\fi                                         
	\fi
}

\flushbottom
\maketitle

\section*{Background \& Summary}
The global drive towards decarbonization causes a transformation of electrical power systems. The rapid integration of distributed renewable energy resources (DERs) increases the operational complexity of the grid and causes significant challenges~\cite{papadisChallengesDecarbonizationEnergy2020}. In particular, power system protection faces a variety of issues due to inverter-based resources (IBRs)~\cite{hooshyarDistanceProtectionLines2015, loebelAnalyticalInvestigationInfluence2025, baeckelandDistanceProtectionPower2022}. To alleviate these challenges, measurement and communication infrastructure is expanded to transform power systems towards smart grids~\cite{dileepSurveySmartGrid2020}. This transformation enhances data availability, particularly through digital substations and the IEC 61850-9-2 Sampled Values (SV) communication protocol. SVs provide synchronized instantaneous, sinusoidal voltage and current measurements called point-on-wave (PoW) data, which are particularly useful to advance the field of power system protection~\cite{aftabIEC61850Based2020}. The development of virtualized protection platforms allows analyzing these measurements using more complex and centralized algorithms~\cite{rubioSmartGridProtection2025}. 

\begin{figure*}[!htpb]
	\centering
	\includegraphics[width=89mm]{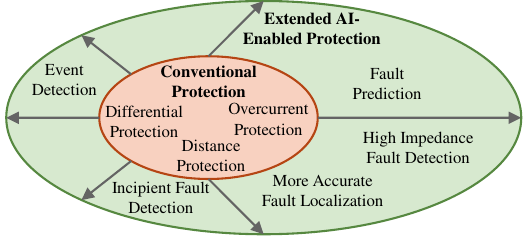}
	\caption{Visualization of developments in AI-Enabled power system protection.}
	\label{fig:2}
\end{figure*}

Therefore, protection strategies are being enhanced by improving conventional protection schemes like distance protection or overcurrent protection and extending protection capabilities to new areas as shown in \autoref{fig:2}. Current research focuses on employing machine learning (ML) based methods ranging from simpler models like random forest or support vector machines~\cite{oelhafSystematicEvaluationMachine2025, muzzammelSupportVectorMachine2020}, to artificial neural networks (ANNs) and deep learning (DL)~\cite{alhanafIntelligentFaultDetection2023, diefenthalerArtificialNeuralNetworks2023}, and graph neural networks (GNNs) that facilitate generalization across different grid topologies\cite{kordowichGraphNeuralNetworkBased2025,freitasFaultLocalizationMethod2021}. These approaches extract waveform features inaccessible to conventional methods, allowing applications like event detection~\cite{wilsonGridEventSignature2024, alacaEventTypeIdentificationPower2024,yangEventDetectionLocalization2023}, improved fault detection, classification and localization~\cite{oelhafScopingReviewMachine2025, oelhafSystematicEvaluationMachine2025}, high impedance fault detection~\cite{ghaderiHighImpedanceFault2017, gaoAdvancingHighImpedance2024, gomesHighSensitivityVegetationHighImpedance2018} and incipient fault detection~\cite{liIncipientFaultDetection2023, luCableIncipientFault2022, changHybridIntelligentApproach2019}.

While significant progress has been made in this area in recent years, one major obstacle that hinders more rapid developments is the lack of open datasets necessary to train, benchmark, and compare different approaches. Imam et al.~\cite{imamParametricNonparametricMachine2024} identify the lack of datasets as the most significant challenge that prevents the selection of optimal ML models. A recent review paper also highlights the gap in datasets and mentions that only \qty{1.7}{\percent} out of 119 reviewed papers published their data~\cite{oelhafScopingReviewMachine2025}.

Training ML models requires a large and diverse database. In the field of power system protection, previous papers used simulation data, real-world data, or a combination of both. Ideally, ML models would be trained on real-world data to avoid the risks of domain shift caused by real-world noise and sub-optimal simulation when transitioning from the simulation domain to real-world applications. However, Evdakov et al.~\cite{evdakovDatasetRealworldOscillograms2026} note that the collection of a sufficient real-world training dataset is practically impossible as faults and events in power systems are rare and manual labeling is expensive and error-prone. Therefore, a combination of simulation and real-world data is sensible, where the utilization of synthetic data is beneficial for developing ML algorithms as they allow for controlled experiments and balanced datasets~\cite{evdakovDatasetRealworldOscillograms2026}. For simulation datasets, Oelhaf et al.~\cite{oelhafScopingReviewMachine2025} specify the following key requirements: (i) diverse grid topologies, fault scenarios and operating conditions; (ii) raw signal export for further processing; (iii) comprehensive labeling and documentation.

Several real-world datasets exist that can complement our proposed datasets. Evdakov et al.~\cite{evdakovDatasetRealworldOscillograms2026} provide 50,000 partially labeled real-world oscillograms. The Grid Event Signature Library~\cite{wilsonGridEventSignature2024} is a smaller but fully labeled dataset, while Réseau de Transport d'Electricité (RTE) provides additional unlabeled voltage and current waveforms~\cite{DatabaseRTE}.

For synthetic data, multiple studies contributed datasets for steady-state load flow studies\cite{krishnanValidationSyntheticUS2020a,en13123290,schweitzerAutomatedGenerationAlgorithm2017a}. However, these datasets target a different set of problems, which require accurate representation of daily load flows by creating synthetic time-series data with minute-level resolution. In contrast, we focus on transient phenomena, where modeling in the range of milliseconds is crucial. Additional studies often generate time-series datasets with reduced temporal resolution in the phasor domain (non-point-on-wave)~\cite{zhengMultiscaleTimeseriesDataset2022,dabouBigDataModelingBased2025}. However, phasor domain simulations cannot capture transients from many events relevant for power system protection and phasor domain data do not allow sub-cycle analysis, limiting the speed of fault detection algorithms. Other studies created synthetic waveform data tailored to specific use cases, such as high-impedance fault (HIF) detection\cite{yangIEEE34Nodes2024} or transformer fault analysis~\cite{beraTransientsFaultsTransformer2020}. Machlev et al.~\cite{machlevOpenSourceDataset2021} provide a dataset that does not rely on simulation and therefore limits its ability to capture the root cause of disturbances in power systems. The datasets by Ross and Mahapatra\cite{rossIRTSDOpenSourceData} as well as by Kouraichi et al.\cite{kouraichiComprehensiveDatasetSimulation2026} are the most similar to our approach but contain data from only one single grid topology. Our previous dataset also covers only one single topology with less detailed modeling\cite{oelhafPROTECT90FaultDataset2026}. Therefore, to our knowledge, no existing public dataset satisfies the criteria outlined by Oelhaf et al.~\cite{oelhafScopingReviewMachine2025}. 

We want to fill this gap by creating a comprehensive open-source dataset called EvEMTBench, that covers the area of advanced power system protection by simulating faults and events in electromagnetic transient (EMT) simulations. Specifically, we want to cover the use cases of: (i) general-purpose event detection (e.g., Alaca et al.\cite{alacaEventTypeIdentificationPower2024}); (ii) fault detection, classification and localization (e.g. Oelhaf et al.\cite{oelhafControlledComparisonMachine2025}); (iii) incipient fault detection (e.g. Li et al.\cite{liIncipientFaultDetection2023}). EvEMTBench's goal is to cover training, fine-tuning, and benchmarking by including a diverse set of grids, voltage levels and events. We generate the dataset by automatically executing a large number of simulations in DIgSILENT PowerFactory using detailed EMT models. The grid models contain both conventional power plants using synchronous machines as well as IBR-based generation. The EvEMTBench dataset contains comprehensively labeled synchronized grid-wide measurements of PoW data with a frequency of \qty{9600}{\hertz} of a wide variety of events in power systems. We publish the data on FAUDataCloud\cite{zotero-item-36862}.

\section*{Methods}\setcurrentname{Methods}
\label{sec:methods}
The development of ML models for power system protection requires high-quality training data that accurately represents grid behavior during faults and events. We incorporate both 50 Hz and 60 Hz grids in the EvEMTBench dataset, though with primary emphasis on European grid configurations. The dataset is designed to serve as a foundation for training, validating, and benchmarking ML models across multiple advanced protection tasks. Previous studies have shown that training on simulation data can be useful when deploying ML models on real-world data\cite{gaoAdvancingHighImpedance2024}. However, the main purpose of this dataset is to provide a basis for fair benchmarks of different approaches and ML models, as performance may deteriorate in real-world applications due to the simulation-to-real gap, often requiring domain adaptation or fine-tuning with grid-specific data.

The realism of synthetic datasets critically depends on appropriate model selection and parameterization. Therefore, we provide a very detailed modeling description in the following sections to foster confidence in the dataset. To ensure physical plausibility, we conducted an extensive literature review to establish realistic parameter ranges for all power system components. If not noted otherwise, we use random sampling from uniform distributions, because the real distribution of grid parameters is generally unknown\cite{erlinghagen2019elektromechanische}. This choice is justified because uniform distributions are suitable for generating a diverse set of power system simulations\cite{erlinghagen2019elektromechanische}.

 The remainder of this section details our methodology in the following order: first, we describe the simulation framework and an overview of the data generation process. Second, we introduce the different types of datasets we create. Subsequently, we describe the selected grid models and explain the simulated events and associated parameter distributions. Finally, we provide a detailed description of the component parameterization procedure, along with tables summarizing realistic parameter ranges of randomly sampled parameters for all relevant grid elements.

\subsection*{Overview of the Simulation Framework}\setcurrentname{Overview of the Simulation Framework}
\label{sec:overview}
\begin{figure}
	\centering
	\includegraphics[width=182mm]{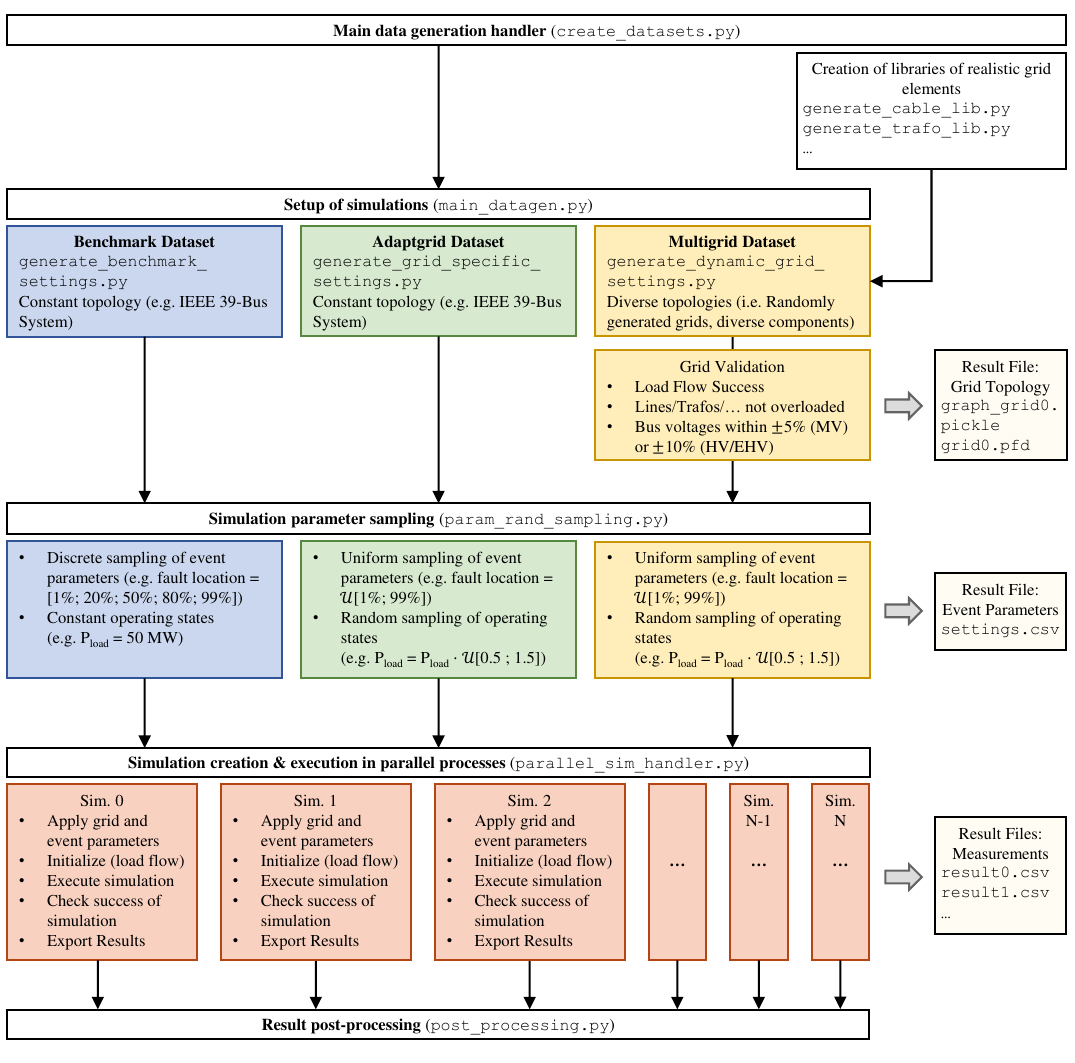}
	\caption{Visualization of main framework.}
	\label{fig:framework}
\end{figure}
To generate the EvEMTBench dataset, we developed the framework shown in \autoref{fig:framework}. Within this framework, DIgSILENT PowerFactory is utilized as the core simulation engine. PowerFactory can be automated via its Python API and accurately represents transient processes in EMT simulations while generally showing good agreement with other EMT tools like PSCAD~\cite{heisselComparativeEMTsimulationsEnergisation2019,lund2005dynamic,ganeshanModelValidationThreePhase2020}.

In EMT simulation-based dataset generation, a fundamental trade-off exists between model accuracy and practical implementation. While theoretically more complex models could better approximate reality, they become increasingly computationally expensive as well as difficult to parameterize with values available in literature. Additionally, EMT models often require different modeling approaches depending on the considered frequency ranges. Therefore, following established guidelines for network element representation during transient analysis, we focused on accurately modeling Group I frequencies (0.1 Hz to 3 kHz)\cite{GuidelinesRepresentationNetwork1990}, which includes the most relevant transient phenomena for extended protection applications.

The EvEMTBench dataset generation process starts by generating libraries of realistic grid elements, like transformers, synchronous machines, or cables. These elements are subsequently used in automatically generated grid models as described in Section \nameref{sec:grids}. After generating these libraries, simulations for the three dataset types described in Section \nameref{sec:datasettypes} are configured. For each power system model, its topology is stored as a machine-readable NetworkX graph serialization in a file called \texttt{graph\_gridN.pickle} where relevant connections and parameters of grid elements like buses, overhead lines (OHLs), or synchronous machines can be accessed. Here, the N stands for the grid number. We use NetworkX Version 3.2.1 and Python 3.9 for serialization. Additionally, we export the basic topology as a proprietary PowerFactory \texttt{.pfd} file to enable importing the grid directly into PowerFactory for in-depth analysis. These files are the only ones in a proprietary format, serve as complementary information and are not required for using the dataset.

For each simulation, a set of event parameters and operating states is sampled. These parameters are stored in a file named \texttt{settings.csv}, which includes event parameters such as fault resistance, as well as operating state parameters like the active power demand of loads. After the parameter sampling is complete, a large number of independent EMT simulations is executed in parallel processes. For each simulation, the set of previously determined parameters is applied in the PowerFactory model. The simulation is then initialized using an unbalanced load flow, followed by the execution of the EMT simulation itself. 

The EMT simulation uses a step size of \qty{10}{\micro\second}, which is a generally accepted step size for EMT simulations. The simulation length is \qty{1.5}{\second}. We discard the first \qty{1.0}{\second} to allow settling of possible initial transients from suboptimal initialization procedures. We export voltages and currents at each cubicle of the main voltage level of the grid (e.g., the cubicles marked in black in \autoref{fig:framework}) for the last \qty{0.5}{\second} of the simulation. The export frequency is \qty{9600}{\hertz}, which captures the relevant signal frequencies of Group I transients and is suitable for both \qty{50}{\hertz} and \qty{60}{\hertz} grids. More specifically, the measurement data tensor $\mathbf{M}$ has the following form (both window endpoints are included):
\begin{equation}
	\mathbf{M} \in \mathbb{R}^{n \times c \times T},
	\quad
	\text{where }
	n = n_{\mathrm{cubicles}},
	\quad
	c = [v_{\mathrm{a}}(t), v_{\mathrm{b}}(t), v_{\mathrm{c}}(t), i_{\mathrm{a}}(t), i_{\mathrm{b}}(t), i_{\mathrm{c}}(t)], 
	\quad
	T = \qty{0.5}{\second} \cdot \qty{9600}{\hertz} + 1 = 4801
	\label{eq:measurement_tensor}
\end{equation}
Here, $v_{\mathrm{a}}, v_{\mathrm{b}}, v_{\mathrm{c}}$ are the phase to ground voltages, and $i_{\mathrm{a}}, i_{\mathrm{b}}, i_{\mathrm{c}}$ are the phase currents.

The data is only exported if both initialization and simulation are successful. Therefore, the total number of usable data records is slightly less than 10000 simulations per dataset depending on how many simulations failed. Parameter sets that do not lead to converging load flows or lead to numerical instabilities during the EMT simulation are discarded. These parameter sets are also removed from the settings file during post-processing of the dataset and the final labels are stored in a file called \texttt{settings\_clean.csv}. The data is exported to one result file per simulation named \texttt{result\{N\}.csv}. Here, \texttt{N} stands for the simulation number, which can be used to map the parameters defined in \texttt{settings\_clean.csv} to the exported transient characteristics. The combination of result files and setting files forms the core of the datasets that can be used for training and benchmarking of ML classifiers for advanced power system protection tasks.

\subsection*{Dataset Types}\setcurrentname{Dataset Types}
\label{sec:datasettypes}
Previous papers often simulate one single grid and subsequently split the data into a train and test dataset. While this is useful for initial experiments, results on these grids do not necessarily translate to real-world use cases, as they implicitly assume ideal grid models. This assumption may not hold in real-world scenarios, as grid models often have parameter imperfections, and models of connected elements such as consumers or IBR controllers are typically unknown. Additionally, models trained for specific grids must be retrained for each new grid, which hinders scalability.

To alleviate this challenge, previous research has utilized the concept of \textit{domain randomization}, where relevant parameters of the simulation are systematically varied to introduce enough variability in the simulation that real-world data appear as one instance within the distribution of simulated scenarios\cite{tobinDomainRandomizationTransferring2017}. In general, three classes of parameters can be randomly sampled: (i) event parameters, like fault resistance or fault location, (ii) operating states, like load demand, or active power of synchronous machines, (iii) grid level parameters, like grid topology, line lengths or placement of grid connected converters. By creating three different types of datasets, which are shown as blue, green and yellow in \autoref{fig:framework}, we aim to enable a cross-topology domain-generalization with an out-of-distribution evaluation protocol.

As our goal is to create a dataset that can be used to demonstrate cross-grid generalization, we vary all three classes of parameters to create the \textit{``multigrid dataset"}. For this purpose, we create simulations that consist of a set of randomly sampled, yet realistic grids of different voltage levels. Each grid has a unique topology and contains a variety of elements (e.g. overhead lines, transformers, etc.). The realism of each grid is ensured by parameterizing its elements with realistic values, as detailed in the subsequent subsections. Additionally, we verify load flow convergence and check for maximum overload conditions of 110\% for transformers, lines, and synchronous machines. We also ensure that voltage deviations remain between \qty{0.95}{\pu} and \qty{1.05}{\pu} for MV grids, and \qty{0.90}{\pu} and \qty{1.10}{\pu} for HV and EHV grids. In each grid that fulfills these requirements, we simulate commonly occurring faults and operating events described in Section \nameref{sec:events} with randomly sampled parameters. The resulting database can be used to train machine learning models for advanced protection tasks. In total, we simulate 10000 events in 105 topologies per voltage level.

To evaluate and benchmark the trained machine learning models, a \textit{``benchmark"} dataset is generated for four grids from literature, as described in Section \nameref{sec:grids}. Using established grids from prior work enhances reproducibility and confidence in the dataset. The topology and operating conditions of these grids remain constant across the benchmark dataset. Unlike the training dataset, where event parameters are randomly sampled, the benchmark dataset systematically varies event parameters at discrete, predefined values (e.g., fault location = \{1\%, 20\%, 50\%, 80\%, 99\%\}). This approach allows a systematic evaluation of model performance.

Finally, it is common practice in ML to fine-tune pre-trained models to specific use cases. Therefore, for each grid in the benchmark dataset, we create a complementary \textit{``adaptgrid"} dataset, which can be used to adapt or to train ML models from scratch for specific grids. In the adaptgrid dataset, we simulate randomly sampled events for different operating states of the grid. For this purpose, we assume approximate knowledge of load levels in the grid by scaling the original loads in the grid between \qty{50}{\percent} and \qty{150}{\percent}. This adaptgrid dataset serves as an intermediate between the benchmark and multigrid datasets: while both events and operating states vary, the dataset is generated for a fixed topology instead of randomly sampled topologies. Therefore, this dataset can be used to fine-tune or train models for specific grids. In total, we simulate 10000 events in each of the selected grids from literature.

\subsection*{Grid Models}\setcurrentname{Grid Models}
\label{sec:grids}
\begin{figure*}
	\centering
	\includegraphics[width=1\linewidth]{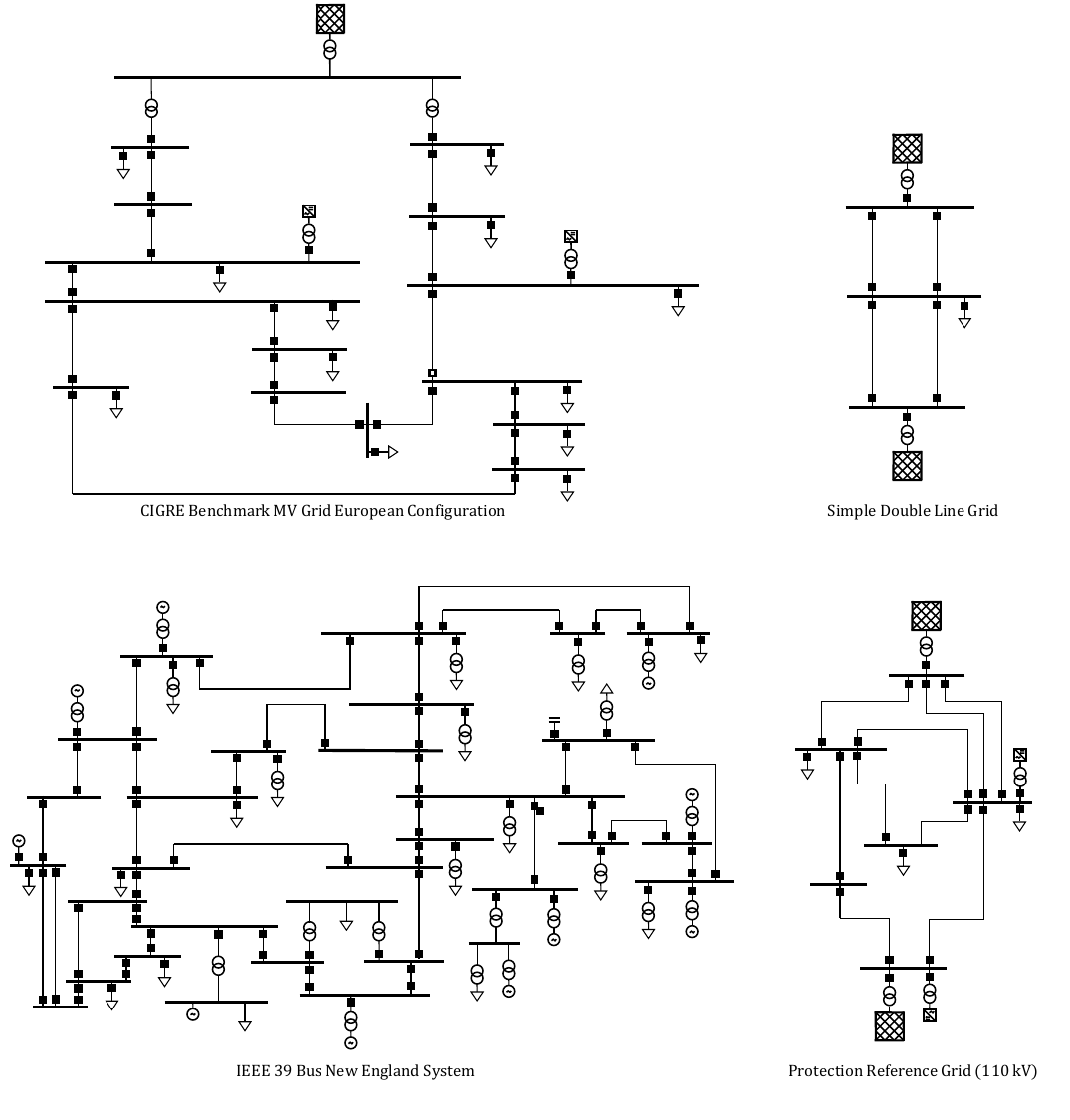}
	\caption{Single line diagrams of considered grids.}
	\label{fig:grids}
\end{figure*}
To enable a fair comparison of machine learning models, we simulate events in four different grids commonly used in literature for the EvEMTBench dataset. A requirement for the selected grids is adequate modeling depth for EMT-simulations. Therefore, some widely used grids, like the IEEE 9-bus system, are excluded. The grids shown in \autoref{fig:grids} were selected to test the applicability of machine learning algorithms across diverse grid types and voltage levels:

\begin{itemize}
	\item \textbf{CIGRE Benchmark MV Grid European Configuration}\cite{BenchmarkSystemsNetwork2014} (\qty{20}{\kilo\volt}): A relatively small benchmark system proposed by CIGRE, which can be utilized to research the challenges and benefits of advanced power system protection tasks in medium voltage grids.
	\item \textbf{Protection Reference Grid}\cite{lorzInterconnectedGridProtection2023} (\qty{110}{\kilo\volt}): A reference grid developed specifically to demonstrate challenges of power system protection. It can be used to demonstrate ML models' performances in subtransmission grids under IBR impact.
	\item \textbf{IEEE 39 Bus New England System}\cite{4113518} (\qty{345}{\kilo\volt}): One of the most well-known test grids in the area of power systems. Can be used for testing challenges and performances on the transmission level and is the only 60 Hz system in our selection.
	\item \textbf{Simple Double Line Grid}\cite{zieglerDigitalerDistanzschutzGrundlagen2008,oelhafSystematicEvaluationMachine2025} (\qty{110}{\kilo\volt}): A relatively simple grid consisting of four lines and three buses, that can be used for initial benchmarks and proofs of concept. We included it for its simplicity and usability and because it represents crucial standard protection issues like intermediate infeed and influence from parallel lines.
\end{itemize}

Both the CIGRE Benchmark MV Grid European Configuration and the IEEE 39 Bus New England System are strongly based on the implementation examples provided by PowerFactory\cite{MV_Microgrid,39_Bus_New_England_System}.

\subsection*{Events}\setcurrentname{Events}
\label{sec:events}
We simulate a variety of events identified from relevant studies on advanced power system protection\cite{oelhafSystematicEvaluationMachine2025, liIncipientFaultDetection2023, luCableIncipientFault2022, kordowichFeatureSelectionFault2026}. The complete list of events is shown in \autoref{tab:events}. The event start time and fault location are randomly sampled from uniform distributions. The events can be categorized into two types, namely fault events and operating events. Normal operating events include inrushes of transformers connected to higher and lower voltage levels, load switches, switches of capacitors, switches of overhead lines and cables as well as starting induction machines in medium voltage systems. We do not simulate motor starts in high-voltage systems, as their impact is less significant in these grids. Additionally, we simulate loss of generation, including both conventional synchronous generators and IBRs.

Additionally, we simulate a wide variety of faults. This includes standard short circuits with constant impedance, including single line to ground (LG), line to line (LL), double line to ground (LLG) and three phase faults (LLL) with impedances of up to \qty{50}{\ohm}\cite{FNN2009}. Additionally, we simulate high impedance ground faults, which are primarily a challenge in medium voltage grids with up to \qty{150}{\kilo\ohm}. The HIF resistance is sampled in four discrete ranges representing different impedance levels: [\qtyrange{50}{500}{\ohm}], [\qtyrange{500}{5000}{\ohm}], [\qtyrange{5000}{50000}{\ohm}], and [\qtyrange{50000}{150000}{\ohm}], with uniform probability within each range and equal probability between ranges. Additionally, we simulate incipient faults, which are usually single phase faults with high impedances that extinguish within less than four cycles\cite{zhangMulticycleIncipientFault2017}. Single line to ground faults often exhibit arc characteristics which we represent using Kizilcay's arc model\cite{kizilcayNumericalFaultArc1994}. This arc modeling approach is described in more detail in Section \nameref{sec:arcs}.

The event parameters are summarized in \autoref{tab:event_parameters}.

\begin{table}
	\centering
	\caption{List of simulated events.}
	\label{tab:simulated-events}
	\begin{tabular}{%
			l 
			>{\raggedright\arraybackslash}p{0.4\linewidth} 
			c 
			c 
			c 
		}
		\toprule
		\textbf{Event name} & \textbf{Event description} & \textbf{MV} & \textbf{HV} & \textbf{EHV} \\ 
		\midrule
		Load On & Closing of a load-switch connecting a load & Yes & Yes & Yes \\
		Load Off & Opening of a load-switch disconnecting a load & Yes & Yes & Yes \\
		Capacitor On & Closing of a switch connecting a shunt capacitor & Yes & Yes & Yes \\
		Capacitor Off & Opening of a switch connecting a shunt capacitor & Yes & Yes & Yes \\
		OHL On & Energization of an overhead line segment & Yes & Yes & Yes \\
		OHL Off & De-energization of an overhead line segment & Yes & Yes & Yes \\
		Cable On & Energization of a cable segment & Yes & No & No \\
		Cable Off & De-energization of a cable segment & Yes & No & No \\
		Inrush LV Trafo & Energization of a transformer connecting a lower voltage level & Yes & Yes & Yes \\
		Inrush HV Trafo & Energization of a transformer connecting a higher voltage level & Yes & Yes & No \\
		Motor Start & Starting of an induction motor & Yes & No & No \\
		Syn. Machine Trip & Loss of conventional generation & No & Yes & Yes \\
		IBR Trip & Loss of an IBR based generation  & Yes & Yes & No \\
		\addlinespace[0.2cm]
		1PHG SHC & Single-phase-to-ground short-circuit & Yes & Yes & Yes \\
		1PHG SHC with Arc & Single-phase-to-ground short-circuit with arc & Yes & Yes & Yes \\
		2PH SHC & Two-phase (line-line) short-circuit & Yes & Yes & Yes \\
		2PHG SHC & Two-phase-to-ground short-circuit & Yes & Yes & Yes \\
		3PH SHC & Three-phase short-circuit & Yes & Yes & Yes \\
		1PHG HIF & Single-phase high-impedance ground fault & Yes & No & No \\
		1PHG HIF with Arc &  Single-phase high-impedance ground fault with arc & Yes & No & No \\
		Incipient Fault & Self-extinguishing high-impedance fault & Yes & Yes & Yes \\
		Incipient Fault with Arc & Self-extinguishing high-impedance fault with arc & Yes & Yes & Yes \\ 
		\bottomrule
	\end{tabular}
	\label{tab:events}
\end{table}

\subsection*{Topology}\setcurrentname{Topology}\label{sec:topology}
To avoid overfitting on specific topologies, we generate different synthetic transmission and distribution grids for the multigrid dataset. The goal of these grids is not to represent full scale transmission networks, but instead small sections. The rationale for this is that transients from electrically distant events generally do not influence protection systems, which are therefore relatively local. Similarly, we do not run EMT simulations of full scale distribution networks, but of smaller grids with aggregated lines and loads, which allows the monitoring of the data of all buses without producing prohibitively large amounts of data. The HV and EHV grids have similar generation processes while the MV grids are generated separately to represent the significant differences in structure. The generation process is loosely inspired by the SimBench project\cite{8669482, en13123290}.

For the high voltage grids, we first determine the number of nodes. Subsequently, we generate a random graph with the specified number of nodes and edges, where nodes represent buses and edges represent lines. The number of lines is chosen within a range that ensures a similar average degree as in the SimBench EHV and HV grids\cite{en13123290}. We ensure that no bus is connected to more than five lines and that no bus is isolated by removing and adding lines. Finally, we add two additional lines next to two existing lines to represent parallel lines, a feature commonly found in real-world grids as well as in reference systems like the IEEE 39-Bus system and the 110 kV Test Grid. We do not consider cables in HV and EHV grids, as most transmission and subtransmission circuits are overhead\cite{shortElectricPowerDistribution2004}. To determine the line lengths we analyzed the SimBench EHV/HV grid and used the $5^{\textit{th}}$ and $95^{\textit{th}}$ percentile of their line lengths as the minimum and maximum. On each bus a load, a source or both may be connected. In 345 kV grids, we only use synchronous machines as the source type, in 110 kV grids we also simulate IBR based sources as described in Section \nameref{sec:ibr}. The assumed parameters are summarized in \autoref{tab:topology}.

For MV Grids, we assume one main bus that is fed by a higher voltage level. A randomly sampled number of feeders with a randomly sampled number of buses is connected to this main bus. Each of these feeders can contain a number of laterals to represent the branched structure of distribution grids. The end of two of those feeders can be connected to form a ring structure, which can occur in typical MV grids. The line lengths for the MV grids are based on the values provided in the corresponding SimBench paper \cite{8669482}. As the focus in our dataset is on accurate representation of static network characteristics and load flow behavior, compared to their approach, we aggregate buses approximately by a factor of five to avoid producing prohibitive amounts of data. Therefore, the respective line lengths are increased by a factor of five to preserve relevant grid properties. We use both overhead lines and cables, where the percentage of cables is on average 70\% according to relevant literature\cite{heuckElektrischeEnergieversorgungErzeugung2013, en13123290}. Each bus is assigned a load, and an IBR model is connected to approximately \qty{20}{\percent} of buses to represent DERs like solar power plants, which typically feed into MV grids.

The detailed models for all grid elements like OHLs, IBRs, or transformers are randomly chosen from the previously created libraries of randomly sampled grid elements. The parameter ranges of these models are detailed in the following subsections.

A summary of all randomly sampled parameters for synthetic grid topology generation is given in \autoref{tab:topology}. Here, and in the following tables, square brackets indicate uniform sampling between minimum and maximum, while curly brackets indicate discrete sampling.

\subsection*{Modeling of Grid Equivalents}
To model equivalents of adjacent or overlaying grids that typically serve as the slack bus in the simulation, PowerFactory uses ``External Grid" elements. The relevant input parameters of these external grids are listed in \autoref{tab:external_grids} and must be entered in PowerFactory. Typical values for short circuit power could be obtained from relevant literature\cite{k.malekianf.safargholik.kuechm.domagkj.meyerandm.hovenCharacteristicParametersReference2017,heuckElektrischeEnergieversorgungErzeugung2013}. The $R/X$ and $R_0/X_0$ ratios in high voltage and extra high voltage grids are typically in a range between 0.1 and 0.2\cite{heuckElektrischeEnergieversorgungErzeugung2013, heroldElektrischeEnergieversorgung32008}. We assume that $X_0/X_1 \approx Z_0/Z_1$ since $R_0 << X_0$ and utilize the value range of $ Z_0/Z_1$ between 2 and 4 from relevant literature\cite{heroldElektrischeEnergieversorgung32008,doeringVerfahrenZurGerichteten2021}.

We carefully adjust the voltage setpoint to avoid idealized assumptions while ensuring stability and realistic voltage levels in long downstream feeders. The initial phase angle serves as a reference angle for the whole grid model and is varied across the full range of values to avoid aligning the exported voltages to the start of the simulation.

\subsection*{Modeling of Loads}
We model loads based on the widely used, standard polynomial load model also called ZIP load model established in literature\cite{milanoPowerSystemModelling2010}. The active, reactive and apparent power are compiled based on the SimBench grid models for MV, HV, and EHV grids\cite{en13123290}. Typical power factors in distribution grids are given by Short\cite{shortElectricPowerDistribution2004}. The ZIP load parameters are summarized from Erlinghagen\cite{erlinghagen2019elektromechanische}. All relevant parameters aggregated from literature are summarized in \autoref{tab:loads}.

\subsection*{Modeling of Cables}
To ensure an accurate representation of the dynamic behavior of cables, we use a distributed parameter model. More specifically, we utilize PowerFactory's implementation of the universal line model (ULM), also called Gustavsen model\cite{morched1999universal}. 
To enable the utilization of this frequency dependent, distributed line model, a geometric cable model must be utilized. For this purpose, the basic geometry of the cables, as well as conductor and insulation materials and their properties must be entered.

We use single-conductor type cable models, as these are most commonly used for voltages above \qty{10}{\kilo\volt}\cite{shortElectricPowerDistribution2004,heuckElektrischeEnergieversorgungErzeugung2013}. They consist of four layers, namely the conductor, insulation, sheath, and oversheath. Aluminum is predominantly used as the conductor material in modern installations\cite{shortElectricPowerDistribution2004}, while copper is the standard sheath material\cite{shortElectricPowerDistribution2004}.

Three insulation materials are considered in the dataset: polyethylene (PE), cross-linked polyethylene (XLPE), and ethylene propylene rubber (EPR). These materials share identical thickness specifications \cite{shortElectricPowerDistribution2004,oedingElektrischeKraftwerkeUnd2016}, though they differ in their electrical properties as shown in \autoref{tab:cables}. Paper-insulated lead-covered (PILC) cables, which were historically common, are excluded from the model due to their declining use for environmental reasons\cite{shortElectricPowerDistribution2004,schulzElektrischeEnergieversorgung2016}. We consider only the most common oversheath material, namely PE\cite{shortElectricPowerDistribution2004}.

The conductor cross-sectional area ranges from 50 to 500 mm\textsuperscript{2}, covering the most common sizes used in MV networks according to DIN VDE 0276-620 standards\cite{DINVDE0276-620}. The sheath cross-sectional area varies between 16 and 25 mm\textsuperscript{2} depending on the conductor size. The insulation thickness is fixed at 5.5 mm, while the oversheath thickness is fixed at 2.5 mm\cite{glaubitzKabelUndLeitungen1989,shortElectricPowerDistribution2004,oedingElektrischeKraftwerkeUnd2016}.

Cable system layout parameters include burial depth and phase spacing. These parameters were selected based on standard engineering practices and literature values \cite{balzerSchaltUndAusgleichsvorgaenge2016,heroldElektrischeEnergieversorgung22008,oedingElektrischeKraftwerkeUnd2016,BenchmarkSystemsNetwork2014,glaubitzKabelUndLeitungen1989}. Additionally, the earth resistivity must be determined. Generally, the earth resistivity can vary significantly depending on the material in the ground and can reach up to \qty{3000}{\Omega\meter}\cite{balzerSchaltUndAusgleichsvorgaenge2016, heroldElektrischeEnergieversorgung22008}. However, earth resistivities of 5 - 500 \si{\Omega\meter} are more common and are utilized as a plausible range. The electrical parameters incorporate material-specific properties including conductor resistivity, sheath resistivity, and dielectric properties like the relative permittivity $\epsilon_r$ and the dielectric loss factor $\tan\delta$ that vary by insulation type\cite{balzerSchaltUndAusgleichsvorgaenge2016}. Finally, the minimum and maximum frequencies for parameter fitting of the ULM are 0.01 Hz and 1 MHz respectively to ensure the stability of the cable model.

In total, we generate a library consisting of 1000 cables, which are utilized in the multigrid dataset. A summary of all considered parameters and their respective ranges is listed in \autoref{tab:cables}.

\subsection*{Modeling of Overhead Lines}
\begin{figure*}
	\centering
	\includegraphics[width=0.9\linewidth]{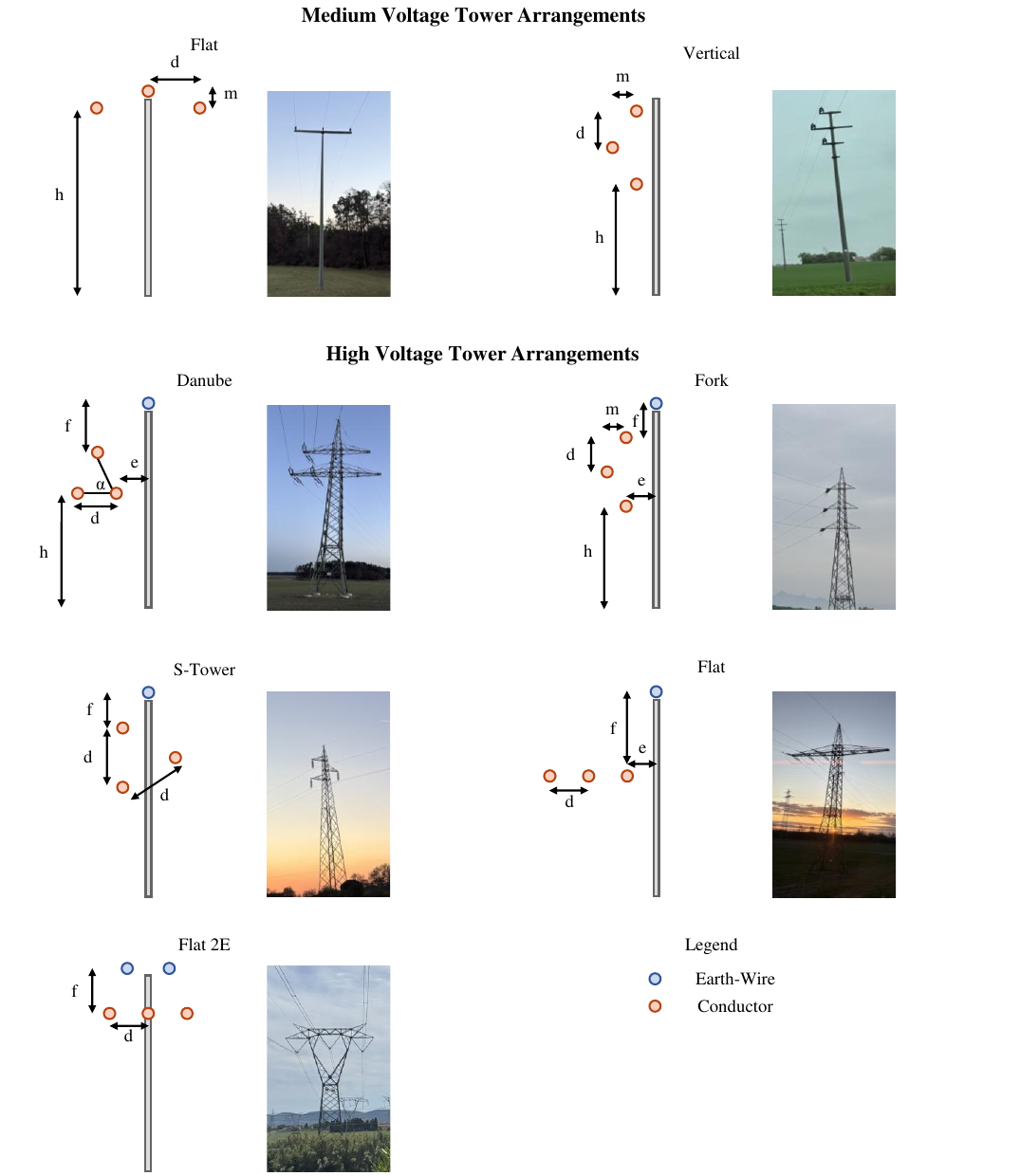}
	\caption{Considered OHL models and photos of real-world equivalents (photographs by the authors).}
	\label{fig:ohlgeo}
\end{figure*}
Similar to the cable modeling, we utilize the ULM as a model for overhead lines. This allows utilizing geometric line models that accurately represent frequency-dependent parameters. These models require a definition of tower geometry and conductor types for each voltage level. The study incorporates multiple tower configurations: For MV towers, we consider flat and vertical arrangements, where conductors are approximately equally spaced in vertical and horizontal direction respectively. For HV towers, we consider Danube-type towers with two conductors side-by-side and one positioned 60 to 90 degrees above, S-towers with equilateral triangular conductor arrangements, Fork towers with approximately vertically stacked conductors, and Flat configurations with either one or two earthwires positioned centrally or laterally. Schematics of the tower configurations used in this study as well as photos from real-world equivalents are shown in \autoref{fig:ohlgeo}.

One simplification we make is that we do not simulate double circuit power lines. For some types two systems are more common than one, but it significantly complicates the data processing and labeling schemes hindering the accessibility of the dataset. Therefore, we follow the approach by Osterkamp et al.\cite{osterkampModellingACDC2025} and utilize the ``half configurations" so that all towers only have one system instead of two as shown in \autoref{fig:ohlgeo}.

Conductor spacing differs significantly across voltage levels. We aggregated typical values from literature~\cite{crastanElektrischeEnergieversorgung12015,flosdorffElektrischeEnergieverteilungMit2008,riegerFreileitungsbau1960} and summarized them in \autoref{tab:ohls}. Tower heights also vary by voltage level, ranging from 9.5~m for 20~kV systems, to 60~m for 380~kV systems according to relevant literature~\cite{crastanElektrischeEnergieversorgung12015,flosdorffElektrischeEnergieverteilungMit2008,riegerFreileitungsbau1960,heuckElektrischeEnergieversorgungErzeugung2013}. From the conductor height at the tower, and the lowest sag, the conductor's average height can be calculated and entered into PowerFactory. The phase order on the towers is randomized to avoid identical configurations for each tower. For high voltage lines we assume the phases are transposed, for medium voltage lines we assume that some lines are transposed and some lines are untransposed. Earth return effects are modeled using Carson's equation. As the OHL is fully determined by geometric parameters, we added the resulting, internally calculated electrical parameters to \autoref{tab:ohls} as well.

Shunt conductance values are derived from empirical measurements by Fernandes et al.~\cite{fernandesTransmissionLineShunt2004}. Finally, identical to the cable models, the minimum and maximum frequencies for parameter fitting of the ULM are 0.01 Hz and 1 MHz respectively.

In addition to the basic geometry of overhead lines, the utilized conductors must be defined. These conductors are selected based on voltage level and operational requirements. For 20 kV systems, we selected conductors with continuous current-carrying capacity between 100 and 650 A\cite{shortElectricPowerDistribution2004}. Aluminum (AL), Aldrey (ALD), and aluminum-steel (AST) conductors are used at this voltage level.

At higher voltage levels, AST conductors are used for almost all HV lines~\cite{heroldElektrischeEnergieversorgung22008}. Therefore, for 110 kV systems, we use AST conductors with cross-sectional areas between 185 and 525~\unit{\milli\meter\squared} according to relevant literature\cite{DIN609090VDE,balzerKurzschlussstromeDrehstromnetzen2020,oedingElektrischeKraftwerkeUnd2016}. At this voltage level, bundle conductors with two conductors or one conductor can be used. At 345 kV, AST conductors with cross-sectional areas ranging from 185 to 800~\unit{\milli\meter\squared} are employed \cite{DIN609090VDE,oedingElektrischeKraftwerkeUnd2016,balzerKurzschlussstromeDrehstromnetzen2020,heroldElektrischeEnergieversorgung22008}. This voltage level uses bundle conductors with three or four conductors, where the optimal conductor spacing ranges from \qtyrange{40}{50}{\centi\meter}\cite{oedingElektrischeKraftwerkeUnd2016}. AST conductors are most commonly used as earth wires with cross-sectional areas from \qtyrange{35}{95}{\milli\meter\squared} according to relevant literature\cite{oedingElektrischeKraftwerkeUnd2016}.

In addition to the previously mentioned parameters, the geometric mean radius (GMR) must be entered into the conductor model of PowerFactory. For solid conductors the GMR can be calculated as:
\begin{equation}
	\text{GMR} = r \cdot e^{-\frac{1}{4}},
\end{equation}
where $ r $ is the conductor radius.

For AST conductors, the GMR is calculated as:
\begin{equation}
	\text{GMR} = r \cdot \exp\left(\frac{3q^2 - r^2}{4(r^2 - q^2)} - \frac{q^4}{(r^2 - q^2)^2} \ln\left(\frac{r}{q}\right)\right),
\end{equation}
where $r$ is the outer radius of the conductor and $q$ is the inner radius of the conductor, which is equivalent to the radius of the steel core.

\autoref{tab:conductors} summarizes the parameter ranges for different conductor types, including nominal cross-section, strand and conductor diameters, DC resistance, and continuous current-carrying capacity. \autoref{tab:cond_select} summarizes the conductor selection criteria for each voltage level. In total, we create a library consisting of 1000 OHL types per voltage level and nominal frequency.

\subsection*{Modeling of Transformers}

Transformers are one of the most important elements of electric grids and therefore modeled in great detail. The detailed modeling approach is especially relevant for the accuracy of transients from transformer inrush events. Therefore, common rated transformer powers $S_{\mathrm{n}}$, along with their corresponding short circuit voltages $u_{\mathrm{k}}$, winding losses (also called copper losses) $P_{\mathrm{k}}$, no load losses (also called iron losses) $P_{\mathrm{l}}$, as well as the no load current $I_{\mathrm{l}}$ are obtained from literature\cite{heroldElektrischeEnergieversorgung22008, roeperKurzschlussstroemeDrehstromnetzen1984, k.malekianf.safargholik.kuechm.domagkj.meyerandm.hovenCharacteristicParametersReference2017, oedingElektrischeKraftwerkeUnd2016, crastanElektrischeEnergieversorgung12015}.

We assume similar ranges for parameters of 50 Hz and 60 Hz transformers since the variance between different 50 Hz transformers mentioned by Herold \cite{heroldElektrischeEnergieversorgung22008} is greater than the variances between 50 and 60 Hz transformers identified by Dawood et al.\cite{dawoodExperimentalAnalysisEffect2025}. One of the most common vector groups of transformers is YNd5/11\cite{crastanElektrischeEnergieversorgung12015}, where YN is at the higher voltage level to reduce insulation requirements. Notable exceptions are the grid coupling transformers between EHV and HV which are more commonly YNyn transformers\cite{crastanElektrischeEnergieversorgung12015} as well as the transformers connecting low voltage loads which are more typically Dyn5 or Yzn5 transformers\cite{crastanElektrischeEnergieversorgung12015}.

Most generated transformers connect two discrete voltage levels per transformer class as shown in \autoref{tab:trafo}. The exception are block transformers connecting generators with different voltages to the primary voltage levels of the transmission and subtransmission systems. These generators typically have voltages of 6.3 kV for units up to 40 MVA, 10.5 kV for those up to 200 MVA, 21 kV for those up to 800 MVA, and 27 kV for those up to 1200 MVA\cite{oedingElektrischeKraftwerkeUnd2016}. 

The zero sequence parameters depend heavily on the vector group and the number of limbs of a transformer. Additionally, YNyn transformers may or may not have internal delta windings that also influence the zero sequence reactance of the transformer. Relevant values from literature\cite{balzerKurzschlussstromeDrehstromnetzen2020, heroldElektrischeEnergieversorgung22008, roeperKurzschlussstroemeDrehstromnetzen1984} are summarized in \autoref{tab:trafo_zero}.

\begin{figure}
	\centering
	\begin{subfigure}{0.4\linewidth}
		\centering
		\includegraphics[width=3.5in]{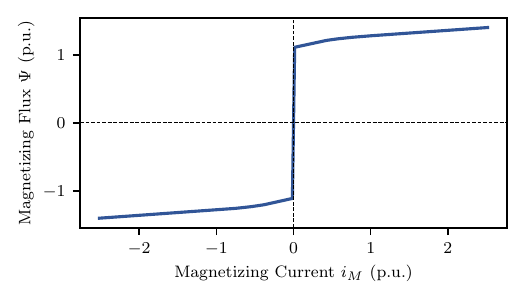}
		\caption{}
		\label{fig:transformersaturation}
	\end{subfigure}
	\hfill
	\begin{subfigure}{0.5\linewidth}
		\centering
		\includegraphics[width=3.5in]{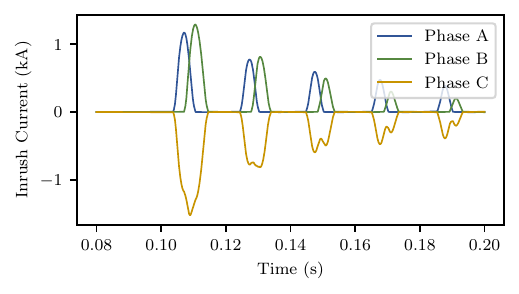}
		\caption{}
		\label{fig:inrushcurrentplot}
	\end{subfigure}
	\caption{Exemplary saturation curve and resulting inrush currents of a transformer.}
	\label{fig:side_by_side}
\end{figure}

Our approach to saturation modeling is mostly in line with the implementation of PowerFactory \cite{digsilentgmbhPowerFactoryTechnicalReference2024}.
The transformer magnetizing characteristic is modeled using a piece-wise function that captures the saturation behavior across three distinct regions:

\begin{equation}
i_{M}(\Psi) =
\begin{cases}
	\operatorname{sign}(\Psi)\left[\dfrac{1}{X_{\mathrm{M,air}}}\left(|\Psi| - \Psi_{\mathrm{knee}}\right) + i_{\mathrm{knee}}\right], & |\Psi| > \Psi_{\mathrm{knee}} \\[10pt]
	\dfrac{1}{X_{\mathrm{M,lin}}}\Psi\left(1 + \left|\dfrac{\Psi}{\Psi_{\mathrm{k0}}}\right|^{k_{\mathrm{sat}}}\right), & 1.11 < |\Psi| \leq \Psi_{\mathrm{knee}}, \\[10pt]
	\dfrac{1}{X_{\mathrm{M,lin}}}\Psi, & |\Psi| \leq 1.11
\end{cases}
\quad \Psi_{\mathrm{k0}} = \Psi_{\mathrm{knee}} \cdot e^{-\frac{1}{k_{\mathrm{sat}}} \ln\left(\frac{\frac{X_{\mathrm{M,lin}}}{X_{\mathrm{M,air}}} - 1}{k_{\mathrm{sat}} + 1}\right)}
\end{equation}
Here, $\Psi_{\mathrm{knee}} = \dfrac{k_{\mathrm{sat}}+1}{k_{\mathrm{sat}}}\Psi_0$ is the knee-point flux, and $i_{\mathrm{knee}} = \dfrac{1}{X_{\mathrm{M,lin}}}\Psi_{\mathrm{knee}}\left(1 + \left|\dfrac{\Psi_{\mathrm{knee}}}{\Psi_{\mathrm{k0}}}\right|^{k_{\mathrm{sat}}}\right)$ is the corresponding knee-point current. $X_{\mathrm{M,lin}}$ and $X_{\mathrm{M,air}}$ are the linear magnetizing reactance and saturated magnetizing reactance respectively. Here, the linear magnetizing reactance corresponds to the reciprocal of the imaginary part of the no-load current\cite{digsilentgmbhPowerFactoryTechnicalReference2024}. The parameter $k_{\mathrm{sat}}$ controls the sharpness of the saturation transition and is chosen as $k_{\mathrm{sat}} = 15$, resulting in an exemplary saturation curve shown in \autoref{fig:transformersaturation}. One example of resulting three phase inrush currents is shown in \autoref{fig:inrushcurrentplot}. The relevant parameters for the saturation characteristic are listed in \autoref{tab:trafo_saturation}.

In total we generate 1000 types per transformer class, except for block transformers, where we generate 4000 to reflect the different voltage levels that block transformers can be connected to.

\subsection*{Modeling of Synchronous Machines}
For synchronous machines, we utilize PowerFactory's standard model, namely model 2.2 described in the IEEE guide for synchronous generator modeling practice ~\cite{1251520}. It represents synchronous machines using the complete second order model equivalent circuits according to Kundur~\cite{kundurPowerSystemStability1994}. As this model is extensively described in literature, we do not detail its specifics but summarize all relevant parameters in \autoref{tab:syn_mach}. We simulate both salient pole rotors mainly used for hydro power plants and round-rotor synchronous machines mainly used in steam or gas power plants. The saturation of the synchronous machine is represented by an exponential saturation characteristic proposed by Kundur~\cite{kundurPowerSystemStability1994}, where $S_{G1.0}$ and $S_{G1.2}$ define the saturation curve. In total, we generate a library of 1000 different synchronous machine types. The parameters summarized in \autoref{tab:syn_mach} were compiled from relevant literature\cite{andersonPowerSystemControl1982,schulzElektrischeEnergieversorgung2016,balzerSchaltUndAusgleichsvorgaenge2016, balzerKurzschlussstromeDrehstromnetzen2020,heroldElektrischeEnergieversorgung32008, oedingElektrischeKraftwerkeUnd2016}. 

\subsection*{Modeling of Excitation Systems of Synchronous Machines}
Excitation systems are responsible for the voltage control of generators and relevant for the considered transients in the chosen \qty{500}{\milli\second} window in this study due to their relatively fast time constant. In contrast, turbine governors typically have larger time constants and are therefore neglected in this study. In practice, a large number of very diverse excitation systems is used. Therefore, in literature, generic exciter models that aim to capture the characteristics of real-world exciters are used. The IEEE ST1A and the simple exciter system (SEXS) models are commonly used. However, Burlakin et al.\cite{burlakinDynamicModellingTransmission2022} find that their responses are unrealistically fast and recommend the AC1A as a more realistic exciter model. Therefore, we utilize PowerFactory's implementation of the AC1A exciter. The parameters of the AC1A are summarized in \autoref{tab:exciter_ac1a} and follow the IEEE recommended practice for excitation system models\cite{182869}. Each parameter is varied in a range of $\pm$10\% unless specified otherwise to avoid overfitting to specific exciter parameters.

\subsection*{Modeling of IBRs}\setcurrentname{Modeling of IBRs}\label{sec:ibr}
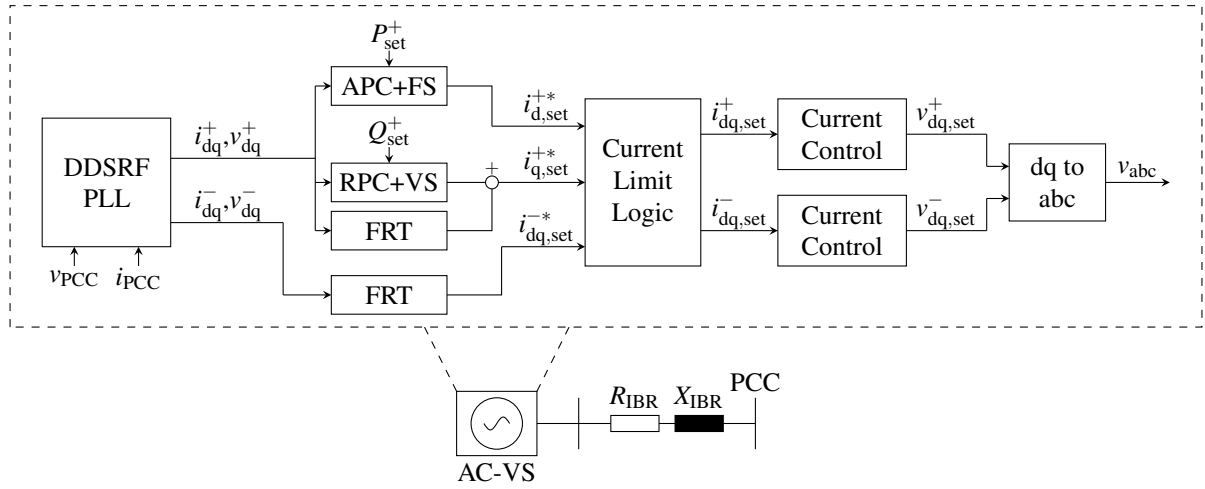
\begin{figure}[t!] 
	\centering
	\begin{tikzpicture}[scale=0.85]
		
		\draw[fill=white, draw=black] (8.20,-0.25) rectangle (6.95,0.75);
		\draw[fill=white, draw=black] (7.575,0.25) circle (0.4);
		\node at (7.575, -0.50) {AC-VS};
		
		\draw (7.775,0.25) sin (7.675,0.35) cos (7.575,0.25) sin (7.475,0.15) cos (7.375,0.25);
		
		\draw (8.85, 0.65) -- (8.85, -0.15);
		\draw (11.60, 0.65) -- (11.60, -0.15);
		
		\node at (11.60, 0.90) {PCC};
		
		\draw (8.20, 0.25) -- (11.60, 0.25);
		
		\draw[fill=white, draw=black] (9.35,0.125) rectangle (10.10,0.375);
		\node at (9.725, 0.65) {$R_\mathrm{IBR}$};
		
		\draw[fill=black, draw=black] (10.35,0.125) rectangle (11.10,0.375);
		\node at (10.725, 0.65) {$X_\mathrm{IBR}$};
		
		\draw[dashed] (0,1.75) -- (0,6.75) -- (18.55,6.75) -- (18.55,1.75) -- (0,1.75);
		\draw[dashed] (6.95,0.75) -- (6.45,1.75);
		\draw[dashed] (8.20,0.75) -- (8.7, 1.75);
		
		\draw (0.5,3) -- (0.5,5) -- (2.5,5) -- (2.5,3) -- (0.5,3);
		\node at (1.5, 4) {\parbox{2cm}{\centering DDSRF\\PLL}};
		\draw[<-, -stealth] (1,2.7) -- ++ (0, 0.3);
		\node at (1, 2.5) {$v_{\mathrm{PCC}}$};
		\draw[<-, -stealth] (2,2.7) -- ++ (0, 0.3);
		\node at (2, 2.5) {$i_{\mathrm{PCC}}$};
		
		\draw[<-, -stealth] (5.9,6.050) -- ++ (0, -0.25);
		\node at (5.9, 6.300) {$P^+_{\mathrm{set}}$};
		\draw[fill=white, draw=black] (5.0,5.200) rectangle ++(1.8,0.6);
		\node at (5.9, 5.500) {APC+FS};
		
		\draw[<-, -stealth] (5.9,4.550) -- ++ (0, -0.25);
		\node at (5.9, 4.800) {$Q^+_{\mathrm{set}}$};
		\draw[fill=white, draw=black] (5.0,3.700) rectangle ++(1.8,0.6);
		\node at (5.9, 4.000) {RPC+VS};
		
		\draw[fill=white, draw=black] (5.0,2.950) rectangle ++(1.8,0.6);
		\node at (5.9, 3.250) {FRT};
		
		\draw[fill=white, draw=black] (5.0,1.950) rectangle ++(1.8,0.6);
		\node at (5.9, 2.250) {FRT};
		
		\draw (2.5, 4.375) -- (4.75, 4.375);
		\draw (4.75, 5.500) -- (4.75, 3.250);
		\draw[<-, -stealth] (4.75,5.500) -- ++ (0.25, 0);
		\draw[<-, -stealth] (4.75,4.000) -- ++ (0.25, 0);
		\draw[<-, -stealth] (4.75,3.250) -- ++ (0.25, 0);
		\node at (3.4, 4.675) {$i^+_{\mathrm{dq}}$,$v^+_{\mathrm{dq}}$};
		
		\draw (2.5, 3.375) -- (4.25, 3.375);
		\draw (4.25, 3.375) -- (4.25, 2.250);
		\draw[<-, -stealth] (4.25,2.250) -- ++ (0.75, 0);
		\node at (3.4, 3.675) {$i^-_{\mathrm{dq}}$,$v^-_{\mathrm{dq}}$};
		
		\draw (6.8, 5.500) -- (7.5, 5.500) -- (7.5, 4.875);
		\draw[<-, -stealth] (7.5,4.875) -- ++ (1.45, 0);
		\node at (8.35, 5.175) {$i^{+*}_{\mathrm{d,set}}$};
		
		\draw (6.8, 4.000) -- (7.5, 4.000);
		\draw (6.8, 3.250) -- (7.5, 3.25) -- (7.5, 4);
		\node at (7.5, 4.225) {\scriptsize$+$};
		\draw[<-, -stealth] (7.5,4) -- ++ (1.45, 0);
		\node at (8.35, 4.3) {$i^{+*}_{\mathrm{q,set}}$};
		\draw[fill=white, draw=black] (7.5,4) circle (0.1);
		
		\draw (6.8, 2.25) -- (7.65, 2.25) -- (7.65, 3);
		\draw[<-, -stealth] (7.65,3) -- ++ (1.3, 0);
		\node at (8.35, 3.3) {$i^{-*}_{\mathrm{dq,set}}$};
		
		\draw[fill=white, draw=black] (8.95,2.7) rectangle ++(1.8,2.6);
		\node at (9.85, 4) {\parbox{2cm}{\centering Current\\Limit\\Logic}};
		
		\draw[<-, -stealth] (10.75,3.25) -- ++ (1.2, 0);
		\node at (11.35, 3.55) {$i^{-}_{\mathrm{dq,set}}$};
		
		\draw[<-, -stealth] (10.75,4.75) -- ++ (1.2, 0);
		\node at (11.35, 5.05) {$i^{+}_{\mathrm{dq,set}}$};
		
		\draw[fill=white, draw=black] (11.95,5.3) rectangle ++(2,-1.1);
		\node at (12.95, 4.75) {\parbox{2cm}{\centering Current\\Control}};
		
		\draw[fill=white, draw=black] (11.95,3.8) rectangle ++(2,-1.1);
		\node at (12.95, 3.25) {\parbox{2cm}{\centering Current\\Control}};
		
		\draw (13.95,3.25) -- ++ (1.25, 0);
		\draw (15.20,3.25) -- ++ (0, 0.5);
		\draw[<-, -stealth] (15.20,3.75) -- ++ (0.35, 0);
		\node at (14.57, 3.55) {$v^{-}_{\mathrm{dq,set}}$};
		
		\draw (13.95,4.75) -- ++ (1.25, 0);
		\draw (15.20,4.75) -- ++ (0, -0.5);
		\draw[<-, -stealth] (15.20,4.25) -- ++ (0.35, 0);
		\node at (14.57, 5.05) {$v^{+}_{\mathrm{dq,set}}$};
		
		\draw[fill=white, draw=black] (15.55,4.6) rectangle ++(1.5,-1.2);
		\node at (16.30, 4) {\parbox{2cm}{\centering dq to\\abc}};
		
		\draw[<-, -stealth] (17.05,4) -- ++ (1, 0);
		\node at (17.55, 4.2) {$v_{\mathrm{abc}}$};

	\end{tikzpicture}
	\caption{GFL IBR model and control scheme}
	\label{fig:IBR}
\end{figure}
DERs like wind and solar power plants are predominantly inverter interfaced. We consider modeling DERs in both MV and HV grids, where they are mostly connected. We assume nominal power of \qtyrange{1}{15}{\mega\VA} in medium voltage grids and \qtyrange{20}{50}{\mega\VA} in high voltage grids\cite{abelsSystemdienlicheAnforderungenDezentrale2023}. The active power setpoint is varied between \qty{0.1}{\pu} and \qty{0.9}{\pu} to represent different levels of solar radiation or wind speed. The IBR is connected via an idealized transformer. All relevant IBR parameters are listed in \autoref{tab:ibr}.

In contrast to synchronous machines, which are predominantly governed by fundamental physical laws, the behavior of IBRs is mostly determined by the inverter control, which is described in the following paragraphs. While grid forming inverters are expected to play a greater role in the future, currently existing IBRs are mostly controlled using grid following (GFL) control algorithms. Therefore, we focus exclusively on GFL IBRs in this study. Similar to excitation systems, a wide variety of IBR controllers can be utilized in practice, motivating the use of a generic model again. The chosen model is shown in \autoref{fig:IBR} and compiled from publicly available literature\cite{Teodorescu_2011,mahrElektrischeEnergiesysteme2021,Sevilmis2019}.

The IBR is modeled as an ideal AC voltage source (AC-VS), connected to the point of common coupling (PCC) via an impedance.
This impedance represents internal IBR impedances, filters and transformers. As we focus on Group I transients as detailed in Section \nameref{sec:overview}, simulating switching of power electronic valves is not necessary and we consider average models based on ideal AC voltage sources.

The current at the PCC is controlled by adjusting the voltage of the AC voltage source. For this purpose, a double decoupled synchronous reference frame (DDSRF) with a phase-locked-loop (PLL) tracks the measured voltage at the PCC and separates the measured phase currents and phase-to-ground voltages at the PCC into positive and negative sequence dq-values. These dq-values are subsequently utilized in the high-level IBR control consisting of an active power control (APC) with frequency support (FS), reactive power control (RPC) with voltage support (VS) and fault-ride-through (FRT) in the positive sequence, as well as a negative sequence FRT component. These controls enable compliance with standard grid codes\cite{IEEE2800_2022,VDE_AR_N_4120_E2024,VDE_AR_N_4110_E2024} both during normal operation and in the event of a grid fault.

During normal operation, the controller supports frequency and voltage stability by adjusting the active power setpoint and reactive power setpoint respectively. During grid faults, faster control is necessary to support grid stability. For this purpose, the positive sequence FRT component supports the grid voltage with capacitive reactive current. The negative sequence FRT complements this mechanism by injecting inductive reactive current in response to unbalanced faults to attenuate the negative sequence voltage and thereby improving the symmetry of the phase voltages.

To consider the thermal limits of the power electronic switching valves and avoid destruction, the IBR currents must be limited\cite{Mozina_2014,Pan_2011}. For this purpose, an unsymmetrical current limitation, which follows relevant literature\cite{Teodorescu_2011,Lee_2011,Winjnhove_2014}, is utilized. If one of the phase currents exceeds the limit of \qty{1.2}{\pu}, the limitation applies the following prioritization: (i) The reactive currents in the positive and negative sequences are assigned equal and highest priority. (ii) The positive-sequence active current is assigned the lowest priority. (iii) The ratio between positive and negative sequence reactive current setpoints is preserved throughout the limiting process.

\subsection*{Modeling of Grounding Systems}
One of the most important factors influencing the characteristics of ground faults and incipient faults is the grounding of the network. Generally the neutral points of the main transformers and grid elements can be either isolated, solidly or effectively grounded using low resistances, resistance grounded or grounded via arc suppression coils (ASC) for resonant grounding. In Europe, compensated grounding is the most common form of grounding for grids between 6~kV and 110~kV\cite{doeringAnalyticalCalculationNeutral2019}. In North America, effectively grounded or resistive grounded networks are more common. For EHV grids, we assume the networks are solidly grounded. Therefore, we simulate the previously mentioned grounding types depending on the voltage level as shown in \autoref{tab:grounding}.

For solidly grounded networks, grounding resistances can vary depending on the grounding quality. We consider the grounding resistance ranges listed in \autoref{tab:grounding} compiled from relevant literature\cite{crastanElektrischeEnergieversorgung12015,shortElectricPowerDistribution2004}. The goal of resistive grounding is to limit short circuit currents of single line to ground faults. The grounding resistance is typically chosen to limit these currents to 1.2 kA - 2 kA in MV grids\cite{ETGFb167Leitfaden2022} and 2 kA - 10 kA in HV grids. This results in the grounding resistances given in \autoref{tab:grounding}.

The goal of compensated networks is to reduce the fault current even further by compensating the capacitive earth current of power lines by tuning an arc suppression coil (ASC). The tuning of the ASC is usually given as $I_{\mathrm{ASC}}$ (\si{\ampere}). This coil tuning current specifies the offset from the grid's total capacitive earth current $I_{\mathrm{CE}}$, where negative values represent undercompensated grids and positive values represent overcompensated grids. Here, overcompensation refers to a state where the ASC current is greater than the capacitive earth current. The reactance of the arc suppression coil can be calculated as follows:
\begin{align}
	I_{\mathrm{CE}} &= \sqrt{3} \cdot \omega \cdot C_{\mathrm{lines}} \cdot V_{\mathrm{n}}\\
	X_{\mathrm{ASC}} &= \frac{V_{\mathrm{n}}}{\sqrt{3} \cdot (I_{\mathrm{CE}} + I_{\mathrm{ASC}})}
\end{align}

Like most real-world grid elements, the ASC is not purely reactive but has a resistive component, which is modeled as a parallel resistance $R_{\mathrm{ASC}}$. The resistive component typically induces currents equal to 2-3\% of the total ASC current \cite{drumlInnovativeMethodenZur,doeringVerfahrenZurGerichteten2021}. The ranges of considered values for all parameters are summarized in \autoref{tab:grounding}.

\subsection*{Modeling of Induction Motors}
PowerFactory has a built-in library of induction motors containing predefined models for various voltage levels. We utilize this library by randomly selecting one of 157 predefined models for each motor starting event.

\subsection*{Modeling of Capacitors}
Capacitors are widely used for reactive power compensation to adjust the power factor and local voltage in the system. Integration of capacitor banks in the power system simulation is especially important as capacitor switching causes relevant transients. We utilize the standard three-phase model for capacitors, where the only relevant parameter is the reactive power. Depending on the voltage level, we aggregated the ranges listed in \autoref{tab:capacitors} from relevant literature~\cite{shortElectricPowerDistribution2004,BenchmarkSystemsNetwork2014,CIGRE_TB_817}. The values are given for all three phases combined.

\subsection*{Modeling of Arc Faults}\setcurrentname{Modeling of Arc Faults}
\label{sec:arcs}
To represent the dynamic processes occurring during arc faults, we chose Kizilcay's arc model\cite{kizilcayNumericalFaultArc1994, kizilcayDigitalSimulationFault1991}. This model describes the arc characteristics using a differential equation and is based on the energy balance in the arc column\cite{kizilcayNumericalFaultArc1994,zhangModelBasedGeneralArcing2016}. The model describes the arc conductance $g$ depending on the arc current $i_{\mathrm{f}}(t)$:
\begin{equation}
	\frac{dg(t)}{dt} = \frac{1}{\tau} \left( \frac{|i_{\mathrm{f}}(t)|}{u_0 + r_0|i_{\mathrm{f}}(t)|} - g(t) \right)
\end{equation}
Here, $g$ is the time-varying arc conductance, $u_0$ is the stationary arc voltage, $r_0$ is the characteristic arc resistance, and $\tau$ is the arc time constant. For secondary arcs, these parameters often fluctuate over time. However, as we only consider primary fault arcs, time constant and conductance can be assumed to be stationary\cite{sousa1995fault, kizilcayDigitalSimulationFault1991}. For 20 kV cables, the arc parameters are given by Zhang et al.\cite{zhangNovelHypothesisTestingBased2025}. For overhead lines, we estimate the time constant $\tau$ from the diagram in Kizilcay's original paper\cite{kizilcayNumericalFaultArc1994}. The arc parameters $u_0$ and $r_0$ depending on the arc length $l_{\mathrm{arc}}$ can be calculated according to Kizilcay and La Seta\cite{kizilcayDigitalSimulationFault2005}:
\begin{align}
	u_0 &= 0.9 \frac{\text{kV}}{\text{m}} \cdot l_{\mathrm{arc}} + 0.4 \text{ kV} \\
	r_0 &= 40 \frac{\text{m}\Omega}{\text{m}} \cdot l_{\mathrm{arc}} + 8 \text{ m}\Omega
\end{align}

\begin{figure}
	\centering
	\includegraphics[width=89mm]{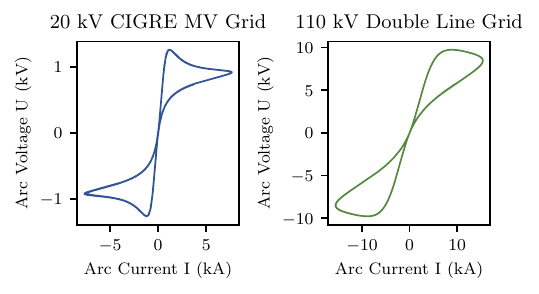}
	\caption{Arc characteristic voltage and current for two different arc parameter sets in a 20 kV and 110 kV grid.}
	\label{fig:arcuiplots}
\end{figure}

Here, we estimate that the arc length can vary between typical insulator lengths and three times a typical insulator length. Typical insulator lengths can be assumed to be \qty{150} - \qty{250}{\milli\meter} for medium voltage grids\cite{heuckElektrischeEnergieversorgungErzeugung2013} and \qty{2}{\meter} and \qty{5}{\meter} for \qty{110}{\kilo\volt} and \qty{380}{\kilo\volt} grids respectively\cite{oedingElektrischeKraftwerkeUnd2016}. These considerations lead to the parameters listed in \autoref{tab:arc_parameters}. Using the model in a power system simulation leads to exemplary arc voltage and current plots shown in \autoref{fig:arcuiplots}. The arc characteristics show general agreement with arc characteristics shown in literature\cite{elkalashyModelingExperimentalVerification2007}.

\subsection*{Parameter Variations}\setcurrentname{Parameter Variations}
\label{sec:parameters}
The following subsection contains tables with all parameters that are randomly sampled during simulations. We hope that these values, compiled from an extensive literature review, can serve as a reference for dataset generation for power system fault and event detection.

\begin{table}[!h]
	\centering
	\caption{Event parameter sampling ranges.}
	\label{tab:event_parameters}
	\begin{threeparttable}
		\begin{tabular}{lc}
			\toprule
			\textbf{Parameter} & \textbf{Sampling Range} \\
			\midrule
			Event/Fault start time (\si{\second}) & [1.1; 1.3]  \\
			Fault location on line (\si{\percent}) & [1; 99]  \\
			Short-circuit resistance (\si{\ohm} ) & [0.001; 50] \\
			High-impedance fault resistance (\si{\ohm}) & [50; 150000]\tnote{a} \\
			Two-phase fault phases & \{a-b, b-c, c-a\} \\
			Single-phase fault phases & \{a, b, c\} \\
			Incipient fault duration (s) & [0.002; 0.08] \\
			\bottomrule
		\end{tabular}
		\begin{tablenotes}
			\item[a] Non-uniform sampling as described in Section \nameref{sec:events}.
		\end{tablenotes}
	\end{threeparttable}
\end{table}

\begin{table*}[!h]
	\centering
	\caption{Network configuration parameters by voltage level.}
	\begin{tabular}{lccc}
		\toprule
		\textbf{Parameter} & \textbf{20 kV} & \textbf{110 kV} & \textbf{345 kV} \\
		\midrule
		MV Number of Feeders & [1; 5] & - & - \\
		MV Number of Nodes per Feeder & [1; 5] & - & - \\
		MV Number of Laterals per Feeder & [0; 2] & - & - \\
		HV/EHV Number of Nodes $N_{\mathrm{nodes}}$ & - & [4; 19] & [4; 19] \\
		HV/EHV Number of Lines & - & [1.2 $\times N_{\mathrm{nodes}}$; 1.6$\times N_{\mathrm{nodes}}$]  & [1.2$\times N_{\mathrm{nodes}}$; 1.6$\times N_{\mathrm{nodes}}$] \\
		HV/EHV Max. Number of Lines per Bus & - & 5 & 5 \\
		Load Connection & True & \{True, False\} & \{True, False\} \\
		Source Connection Prob. (\si{\percent}) & 20 & 50 & 50 \\
		Source Type & IBR & \{Sym, IBR\} & \{Sym\} \\
		Cable Probability & 70\% & 0\% & 0\% \\
		OHL Length (\si{\kilo\meter}) & [0.5; 9.5] & [1; 32] & [1; 111] \\
		Cable Length (\si{\kilo\meter}) & [0.5; 3.75] & - & - \\
		\bottomrule
	\end{tabular}
	\label{tab:topology}
\end{table*}

\begin{table*}[!h]
	\centering
	\caption{Grid equivalent parameters by voltage level.}
	\begin{tabular}{lcc}
		\toprule
		\textbf{Parameter} & \textbf{380/345 kV External Grid} & \textbf{110 kV External Grid} \\
		Grid Equivalent Short Circuit Power $S_{\mathrm{k}}''$ (\si{\mega\volt\ampere}) & [10\,000; 53\,000] & [800; 8\,000] \\
		Resistance-to-reactance ratio $R/X$ & [0.1; 0.2] & [0.1; 0.2] \\
		Zero-sequence to positive-sequence reactance ratio $X_0/X_1$ & [2; 4] & [2; 4] \\
		Zero-sequence resistance-to-reactance ratio $R_0/X_0$ & [0.1; 0.2] & [0.1; 0.2] \\
		Voltage set point $u_{set}$ (pu) & [0.98; 1.02] & [0.98; 1.02] \\
		Initial phase angle $\varphi_{ini}$ (\si{\degree}) & [-180; 180] & [-180; 180] \\
		\bottomrule
	\end{tabular}
	\label{tab:external_grids}
\end{table*}

\begin{table*}[!h]
	\centering
	\caption{Load configuration parameters.}
	\begin{tabular}{lc}
		\toprule
		\textbf{Parameter} & \textbf{Value / Range} \\
		\midrule
		\multicolumn{2}{l}{\textit{Load Model Coefficients}} \\
		\midrule
		$k_{P,Z}$ & [0.20; 1.31] \\
		$k_{P,I}$ & [-1.14; 0.50] \\
		$k_{P,P}$ & [0.14; 0.83] \\
		$k_{Q,Z}$ & [0.43; 4.74] \\
		$k_{Q,I}$ & [-6.19; 0.71] \\
		$k_{Q,P}$ & [-0.14; 2.45] \\
		Normalization constraint & $k_{P,Z} + k_{P,I} + k_{P,P} = 1$ \\
		& $k_{Q,Z} + k_{Q,I} + k_{Q,P} = 1$ \\
		\midrule
		\multicolumn{2}{l}{\textit{Ranges for Active, Reactive and Apparent Power}} \\
		\midrule
		345 kV systems: \\
		\quad Active power $P$ (\si{\mega\watt}) & [9; 326] \\
		\quad Reactive power $Q$ (\si{\mega\VAr}) & [3.6; 116] \\
		110 kV systems: \\
		\quad Active power $P$ (\si{\mega\watt}) & [0; 31] \\
		\quad Reactive power $Q$ (\si{\mega\VAr}) & [0; 12.5] \\
		20 kV systems: \\
		\quad Apparent power $S$ (\si{\mega\VA}) & [0.1; 1.6] \\
		\quad Power factor & [0.8 ind.; 0.95 cap.] \\
		\bottomrule
	\end{tabular}
	\label{tab:loads}
\end{table*}

\begin{table*}[!h]
	\centering
	\caption{Underground cable configuration parameters.}
	\begin{tabular}{lc}
		\toprule
		\textbf{Parameter} & \textbf{Value / Range} \\
		\midrule
		\multicolumn{2}{l}{\textit{Cable Type}} \\
		\midrule
		Rated voltage (\si{\kilo\volt}) & 20 \\
		Insulation materials & \{PE, XLPE, EPR\} \\
		Conductor material & Aluminum \\
		Sheath material & Copper \\
		Conductor cross-sectional area (\si{\milli\meter\squared}) & \{50, 70, 95, 120, 150, 185, 240, 300, 400, 500\} \\
		Sheath cross-sectional area (\si{\milli\meter\squared}) & \{16, 16, 16, 16, 25, 25, 25, 25, 25, 25\} \\
		Conductor radius \textbf{r} (\si{\milli\meter}) & $\sqrt{A_\mathrm{cond}/\pi}$ \\
		Insulation thickness (\si{\milli\meter}) & 5.5 \\
		Sheath thickness (\si{\milli\meter}) & $\sqrt{A_\mathrm{sh}/\pi + (r + t_\mathrm{ins})^2} - (r + t_\mathrm{ins})$ \\
		Oversheath thickness (\si{\milli\meter}) & 2.5 \\
		Conductor resistivity, Al (\si{\micro\ohm\centi\meter}) & 2.8264 \\
		Sheath resistivity, Cu (\si{\micro\ohm\centi\meter}) & 1.7241 \\
		Relative permittivity $\varepsilon_\mathrm{r}$, PE / XLPE / EPR & 2.3 / 2.4 / 3.0 \\
		Dielectric loss factor $\tan\delta$, PE / XLPE / EPR & $0.3 \times 10^{-3}$ / $0.4 \times 10^{-3}$ / $5 \times 10^{-3}$ \\
		\midrule
		\multicolumn{2}{l}{\textit{Cable System Layout}} \\
		\midrule
		Burial depth (\si{\meter}) & [0.7; 1.2] \\
		Cable arrangement & \{Flat, Trefoil\} \\
		Phase spacing, flat (\si{\meter}) & [$d_\mathrm{cab} + 0.07$; 0.3] \\
		Phase spacing, trefoil (\si{\meter}) & $d_\mathrm{cab} + 0.005$ \\
		Earth resistivity $\rho$ (\si{\ohm\meter}) & [5; 500] \\
		\midrule
\multicolumn{2}{l}{\textit{Electrical Parameters (per km)}} \\
\multicolumn{2}{l}{\small \textit{Values represent 5th; \textbf{50th}; 95th percentiles of all randomly generated cables.}} \\
\midrule
Positive Sequence Resistance $R_1'$ (\si{\ohm\per\kilo\meter}) & 0.0841; \textbf{0.2125}; 0.6457 \\
Positive Sequence Reactance $X_1'$ (\si{\ohm\per\kilo\meter}) & 0.1007; \textbf{0.1411}; 0.2688 \\
Positive Sequence Capacitance $C_1'$ (\si{\nano\farad\per\kilo\meter}) & 154.08; \textbf{247.06}; 420.35 \\
Zero Sequence Resistance $R_0'$ (\si{\ohm\per\kilo\meter}) & 0.6561; \textbf{0.7775}; 1.4464 \\
Zero Sequence Reactance $X_0'$ (\si{\ohm\per\kilo\meter}) & 0.2397; \textbf{0.2661}; 0.5100 \\
Zero Sequence Capacitance $C_0'$ (\si{\nano\farad\per\kilo\meter}) & 154.08; \textbf{247.06}; 420.35 \\
		\bottomrule
	\end{tabular}
	\label{tab:cables}
\end{table*}

\begin{table*}[h]
	\centering
	\caption{Overhead line configuration parameters and their respective value ranges.}
	\begin{tabular}{lccc}
		\toprule
		\textbf{Parameter} & \textbf{20 kV} & \textbf{110 kV} & \textbf{345 kV} \\
		\midrule
		Tower arrangements\cite{oedingElektrischeKraftwerkeUnd2016} & \{Flat; Vertical\} & \multicolumn{2}{c}{\{Danube; Fork; S-Tower; Flat; Flat2E\}} \\
		Conductor height at tower \textbf{a} (\si{\meter})\cite{crastanElektrischeEnergieversorgung12015,flosdorffElektrischeEnergieverteilungMit2008,riegerFreileitungsbau1960,heuckElektrischeEnergieversorgungErzeugung2013} & [9.5; 14] & [19; 28] & [22; 60] \\
		Conductor height at lowest sag \textbf{b} (\si{\meter})\cite{schulzElektrischeEnergieversorgung2016} & [6; 0.9$\times$\textbf{a}] & [6; 0.9$\times$\textbf{a}] & [7.8; 0.9$\times$\textbf{a}] \\
		Conductor sag \textbf{c} (\si{\meter}) & \multicolumn{3}{c}{$\textbf{a} - \textbf{b}$} \\
		Conductor average height \textbf{h} (\si{\meter})\cite{oedingElektrischeKraftwerkeUnd2016} & \multicolumn{3}{c}{$\textbf{a} - 2/3 \textbf{c}$} \\
		Conductor spacing \textbf{d} (\si{\meter})\cite{crastanElektrischeEnergieversorgung12015,flosdorffElektrischeEnergieverteilungMit2008,riegerFreileitungsbau1960} & [1.4; 2.2] & [3; 4.7] & [6; 11] \\
		Earthwire horizontal distance to conductor \textbf{e} (\si{\meter})\cite{crastanElektrischeEnergieversorgung12015,flosdorffElektrischeEnergieverteilungMit2008,riegerFreileitungsbau1960} & - & [3.2; 5] & [7; 8] \\
		Earthwire vertical distance to conductor \textbf{f} (\si{\meter})\cite{crastanElektrischeEnergieversorgung12015,flosdorffElektrischeEnergieverteilungMit2008,riegerFreileitungsbau1960} & - & [4; 5] & [8; 15] \\
		Deviation from ideal arrangement \textbf{m} (\si{\meter}) & \multicolumn{3}{c}{[0; 1]} \\
		Danube Tower Conductor Angle \textbf{$\alpha$} (deg) & - & \multicolumn{2}{c}{[60; 90]} \\
		Earth resistivity (\si{\ohm\meter})\cite{balzerSchaltUndAusgleichsvorgaenge2016, heroldElektrischeEnergieversorgung22008} & \multicolumn{3}{c}{[5; 500]} \\
		Shunt conductance (\si{\micro\siemens\per\kilo\meter})\cite{fernandesTransmissionLineShunt2004} & \multicolumn{3}{c}{[1.5e-6; 3.7e-3]} \\
		Bundle spacing (\si{\meter})\cite{oedingElektrischeKraftwerkeUnd2016} & - & - & [0.4; 0.5] \\
		Number of sub-conductors\cite{heuckElektrischeEnergieversorgungErzeugung2013} & 1 & \{1, 2\} & \{3, 4\} \\
		Number of earthwires & 0 & \{1, 2\} & \{1, 2\} \\
		Transposition & \{Yes, No\} & Yes & Yes \\
		\midrule
\multicolumn{4}{l}{\textit{Electrical Parameters (per km)}} \\
\multicolumn{4}{l}{\small \textit{Values represent 5th; \textbf{50th}; 95th percentiles of all randomly generated overhead lines.}} \\
\midrule
Positive Sequence Resistance $R_1'$ (\si{\ohm\per\kilo\meter}) & 0.12; \textbf{0.31}; 1.88 & 0.03; \textbf{0.06}; 0.14 & 0.01; \textbf{0.02}; 0.05 \\
Positive Sequence Reactance $X_1'$ (\si{\ohm\per\kilo\meter}) & 0.35; \textbf{0.39}; 0.44 & 0.25; \textbf{0.29}; 0.40 & 0.23; \textbf{0.26}; 0.29 \\
Positive Sequence Capacitance $C_1$ (\si{\nano\farad\per\kilo\meter}) & 8.24; \textbf{9.44}; 10.47 & 9.08; \textbf{12.58}; 14.62 & 12.49; \textbf{13.82}; 15.70 \\
Zero Sequence Resistance $R_0'$ (\si{\ohm\per\kilo\meter}) & 0.27; \textbf{0.45}; 2.02 & 0.23; \textbf{0.31}; 0.41 & 0.17; \textbf{0.22}; 0.28 \\
Zero Sequence Reactance $X_0'$ (\si{\ohm\per\kilo\meter}) & 1.38; \textbf{1.60}; 1.70 & 0.78; \textbf{1.07}; 1.24 & 0.74; \textbf{0.95}; 1.06 \\
Zero Sequence Capacitance $C_0'$ (\si{\nano\farad\per\kilo\meter}) & 4.09; \textbf{4.45}; 4.83 & 4.70; \textbf{5.59}; 7.03 & 5.53; \textbf{6.32}; 7.72 \\
		\bottomrule
	\end{tabular}
	\label{tab:ohls}
\end{table*}

\begin{table*}[!h]
	\centering
	\caption{Conductor parameters and their respective value ranges.}
	\begin{tabular}{lcccc}
		\toprule
		\textbf{Parameter} & \textbf{AST (Al/St)} & \textbf{AL (Aluminum)} & \textbf{ALD (Aldrey)} & \textbf{Earthwire (Al/St)} \\
		\midrule
		Nominal Cross Section (\si{\milli\meter\squared}) \cite{schulzElektrischeEnergieversorgung2016, crastanElektrischeEnergieversorgung12015} & [16; 1045] & [15.9; 999.7] & [16.0; 999.7] & [35; 95] \\
		Inner Diameter (\si{\milli\meter}) \cite{schulzElektrischeEnergieversorgung2016, crastanElektrischeEnergieversorgung12015} & [1.80; 12.00] & - & - & - \\
		Outer Diameter (\si{\milli\meter}) \cite{schulzElektrischeEnergieversorgung2016, crastanElektrischeEnergieversorgung12015} & [5.40; 43.00] & [5.10; 41.10] & [5.10; 41.10] & [5.40; 13.60] \\
		DC Resistance (\si{\ohm\per\kilo\meter}) \cite{schulzElektrischeEnergieversorgung2016, crastanElektrischeEnergieversorgung12015} & [0.0277; 1.8769] & [0.0336; 2.091] & [0.029; 1.7986] & [0.3060; 0.8342] \\
		Rated Current (\si{\ampere}) \cite{schulzElektrischeEnergieversorgung2016, crastanElektrischeEnergieversorgung12015} & [105; 1580] & [110; 1540] & [105; 1450] & - \\
		\bottomrule
	\end{tabular}
	\label{tab:conductors}
\end{table*}

\begin{table*}[!h]
	\centering
	\caption{Conductor selection criteria by voltage level.}
	\label{tab:cond_select}
	\begin{tabular}{lcc}
		\toprule
		\textbf{Voltage Level} & \textbf{Conductor Type} & \textbf{Selection Criteria} \\
		\midrule
		\multirow{3}{*}{20 \si{\kilo\volt}} & AL & \makecell{100 \si{\ampere} $\leq$ Rated Current $\leq$ 650 \si{\ampere}} \\
		& ALD & \makecell{100 \si{\ampere} $\leq$ Rated Current $\leq$ 650 \si{\ampere}} \\
		& AST & \makecell{100 \si{\ampere} $\leq$ Rated Current $\leq$ 650 \si{\ampere}} \\
		110 \si{\kilo\volt} & AST & \makecell{185 \si{\milli\meter\squared} $\leq$ Nominal Cross Section $\leq$ 525 \si{\milli\meter\squared}} \\
		345 \si{\kilo\volt} & AST & \makecell{185 \si{\milli\meter\squared} $\leq$ Nominal Cross Section $\leq$ 800 \si{\milli\meter\squared}} \\
		Earthwire & AST & \makecell{35 \si{\milli\meter\squared} $\leq$ Nominal Cross Section $\leq$ 95 \si{\milli\meter\squared}} \\
		\bottomrule
	\end{tabular}
\end{table*}

\begin{table}[!h]
	\centering
	\caption{Zero-sequence parameters and limb configuration by transformer type.}
	\begin{tabular}{@{}lcccc@{}}
		\toprule
		\textbf{Type} & \textbf{$\mathbf{S_{\mathrm{n}}}$ Range (MVA)} & \textbf{Limb Type} & \textbf{$\mathbf{x_0/x_1}$ Ratio} & \textbf{$\mathbf{r_0/r_1}$ Ratio} \\
		\midrule
		\multirow{2}{*}{20/0.4} & \multirow{2}{*}{All} & Dyn & [0.95; 1.0] & [1.2; 1.8] \\
		& & Yzn & [0.07; 0.11] & [0.45; 0.54] \\
		\midrule
		110/20 & All & \{3; 5\} & \makecell{[0.85; 1.0] (3-limb)\\{}[0.95; 1.05] (5-limb)} & \makecell{[1.5; 2.8] (3-limb)\\{}[1.05; 1.1] (5-limb)} \\
		\midrule
		110/11 & All & {3; 5} & \makecell{[0.85; 1.0] (3-limb)\\{}[0.95; 1.05] (5-limb)} & \makecell{[1.5; 2.8] (3-limb)\\{}[1.05; 1.1] (5-limb)} \\
		\midrule
		380/110 & All & {3; 5} & \makecell{[1.8; 3.0] (3-limb, delta) \\{} [5.0; 10.0] (3-limb, no delta) \\{} [2.0; 5.0] (5-limb, delta) \\{} [10.0; 100.0] (5-limb, no delta)} & [0.95; 1.05] \\
		\midrule
		345/25 & \textless50 & 3 & [0.85; 1.0] & [1.0; 3.0] \\
		& [50; 350] & 3 & [0.85; 0.95] & [1.8; 2.0] \\
		& \textgreater350 & 5 & [0.95; 1.05] & [1.0; 1.1] \\
		\midrule
		Block & \textless40 & 3 & [0.85; 1.0] & [1.0; 3.0] \\
		& [40; 200] (110 kV) & 3 & [0.85; 1.0] & [1.5; 2.8] \\
		& [40; 200] (345 kV) & 3 & [0.85; 0.95] & [1.8; 2.0] \\
		& \textgreater200 & 5 & [0.95; 1.05] & [1.0; 1.1] \\
		\bottomrule
	\end{tabular}
	\label{tab:trafo_zero}
\end{table}

\begin{table}[!h]
	\centering
	\caption{Important parameters and their respective value ranges of different transformer types.}
	\begin{tabular}{lcccccc}
		\toprule
		\textbf{Class} & \textbf{$\mathbf{S_{\mathrm{n}}}$ (MVA)} & \textbf{$\mathbf{u_{\mathrm{k}}}$ (\%)} & \textbf{$\mathbf{P_{\mathrm{k}}}$ (kW)} & \textbf{$\mathbf{P_{\mathrm{l}}}$ (kW)} & \textbf{$\mathbf{I_{\mathrm{l}}}$ (\%)} & \textbf{Vector Group} \\
		\midrule
		\multirow{5}{*}{20/0.4} &
		0.16 & 4.0 & [2.35; 3.10] & [0.375; 0.460] & [1.1; 1.7] & Dyn, Yzn \\
		& 0.25 & [4.0; 6.0] & [2.75; 4.20] & [0.425; 0.650] & [1.1; 1.7] & Dyn, Yzn \\
		& 0.40 & [4.0; 6.0] & [3.85; 6.00] & [0.610; 0.930] & [0.9; 1.4] & Dyn \\
		& 0.63 & [4.0; 6.0] & [5.40; 8.70] & [0.800; 1.300] & [0.8; 1.2] & Dyn \\
		& 1.00 & 6.0 & [9.50; 13.00] & [1.100; 1.700] & [0.8; 1.2] & Dyn \\
		& 1.60 & 6.0 & [14.00; 20.00] & [1.700; 2.600] & [0.7; 1.1] & Dyn \\
		\midrule
		\multirow{6}{*}{110/20} &
		20 & [10.0; 12.0] & [95; 100] & [14; 16] & [0.4; 0.5] & YNd \\
		& 25 & [10.0; 12.0] & [110; 115] & [16; 19] & [0.4; 0.5] & YNd \\
		& 31.5 & [10.0; 13.0] & [130; 135] & [19; 23] & [0.4; 0.45] & YNd \\
		& 40 & [10.0; 15.0] & [155; 160] & [23; 27] & [0.35; 0.4] & YNd \\
		& 50 & [13.0; 15.8] & 190 & 32 & [0.35; 0.4] & YNd \\
		& 63 & [13.0; 16.0] & 220 & 38 & [0.35; 0.4] & YNd \\
		& 80 & [14.0; 16.0] & 260 & 45 & 0.35 & YNd \\
		\midrule
		110/11 &
		20 & [10.0; 12.0] & [95; 100] & [14; 16] & [0.4; 0.5] & YNd \\
		\midrule
		\multirow{6}{*}{380/110} &
		100 & [14.0; 16.0] & [350; 460] & [40; 60] & [0.35; 0.46] & YNyn \\
		& 160 & [14.0; 16.0] & [480; 560] & [64; 96] & [0.30; 0.35] & YNyn \\
		& 200 & [14.0; 16.0] & [560; 700] & [80; 120] & [0.28; 0.35] & YNyn \\
		& 250 & [15.0; 17.0] & [650; 825] & [100; 150] & [0.26; 0.33] & YNyn \\
		& 300 & [16.0; 19.0] & [690; 930] & [120; 180] & [0.23; 0.31] & YNyn \\
		& 400 & [16.0; 20.0] & [840; 1160] & [160; 240] & [0.21; 0.29] & YNyn \\
		& 500 & [16.0; 20.0] & [1000; 1400] & [200; 300] & [0.20; 0.28] & YNyn \\
		\midrule
		\multirow{7}{*}{345/25} &
		100 & [14.0; 16.0] & [350; 460] & [40; 60] & [0.35; 0.46] & YNd \\
		& 160 & [14.0; 16.0] & [480; 560] & [64; 96] & [0.30; 0.35] & YNd \\
		& 200 & [14.0; 16.0] & [560; 700] & [80; 120] & [0.28; 0.35] & YNd \\
		& 250 & [15.0; 17.0] & [650; 825] & [100; 150] & [0.26; 0.33] & YNd \\
		& 300 & [16.0; 19.0] & [690; 930] & [120; 180] & [0.23; 0.31] & YNd \\
		& 400 & [16.0; 20.0] & [840; 1160] & [160; 240] & [0.21; 0.29] & YNd \\
		& 500 & [16.0; 20.0] & [1000; 1400] & [200; 300] & [0.20; 0.28] & YNd \\
		& 600 & [16.0; 20.0] & [1200; 1680] & [240; 360] & [0.20; 0.28] & YNd \\
		& 800 & [16.0; 20.0] & [1440; 2080] & [320; 480] & [0.18; 0.26] & YNd \\
		\midrule
		\multirow{12}{*}{Block} &
		20 & [10.0; 12.0] & [95; 100] & [14; 16] & [0.4; 0.5] & YNd \\
		& 25 & [10.0; 12.0] & [110; 115] & [16; 19] & [0.4; 0.5] & YNd \\
		& 31.5 & [10.0; 13.0] & [130; 135] & [19; 23] & [0.4; 0.45] & YNd \\
		& 40 & [10.0; 15.0] & [155; 160] & [23; 27] & [0.35; 0.4] & YNd \\
		& 50 & [13.0; 15.8] & 190 & 32 & [0.35; 0.4] & YNd \\
		& 63 & [13.0; 16.0] & 220 & 38 & [0.35; 0.4] & YNd \\
		& 80 & [14.0; 16.0] & 260 & 45 & 0.35 & YNd \\
		& 100 & [14.0; 16.0] & [350; 460] & [40; 60] & [0.35; 0.46] & YNd \\
		& 160 & [14.0; 16.0] & [480; 560] & [64; 96] & [0.30; 0.35] & YNd \\
		& 200 & [14.0; 16.0] & [560; 700] & [80; 120] & [0.28; 0.35] & YNd \\
		& 250 & [15.0; 17.0] & [650; 825] & [100; 150] & [0.26; 0.33] & YNd \\
		& 300 & [16.0; 19.0] & [690; 930] & [120; 180] & [0.23; 0.31] & YNd \\
		& 400 & [16.0; 20.0] & [840; 1160] & [160; 240] & [0.21; 0.29] & YNd \\
		& 500 & [16.0; 20.0] & [1000; 1400] & [200; 300] & [0.20; 0.28] & YNd \\
		& 600 & [16.0; 20.0] & [1200; 1680] & [240; 360] & [0.20; 0.28] & YNd \\
		& 800 & [16.0; 20.0] & [1440; 2080] & [320; 480] & [0.18; 0.26] & YNd \\
		& 1000 & [16.0; 20.0] & [1600; 2400] & [400; 600] & [0.16; 0.24] & YNd \\
		& 1200 & [16.0; 20.0] & [1680; 2640] & [480; 720] & [0.14; 0.22] & YNd \\
		\bottomrule
	\end{tabular}
	\label{tab:trafo}
\end{table}

\begin{table}[!h]
	\centering
	\caption{Transformer saturation parameters.}
	\begin{tabular}{@{}lc@{}}
		\toprule
		\textbf{Parameter} & \textbf{Value} \\
		\midrule
		Flux linkage $\psi_{0}$ (p.u.) & [1.20; 1.25] \\
		Saturated (air core) reactance $X_{\mathrm{M,air}}$ (p.u.) & [1; 2] $\times u_{\mathrm{k}}/100$ \\
		\bottomrule
	\end{tabular}
	\label{tab:trafo_saturation}
\end{table}

\begin{table*}[!h]
	\centering
	\caption{Parameter ranges for synchronous machine types.}
	\begin{threeparttable}
		\begin{tabular}{lcc}
			\toprule
			\textbf{Parameter} & \textbf{Salient Pole (e.g. Hydro)} & \textbf{Round Rotor (e.g. Steam/Gas)} \\
			\midrule
			$S_{\mathrm{gn}}$ (MVA) & [17.5; 700] & [25; 1200] \\
			$\cos \phi_{\mathrm{n}}$ & [0.8; 0.95] & [0.8; 0.975] \\
			$V_{\mathrm{gn}}$ (kV) & \multicolumn{2}{c}{6.3 (if $S_{gn} \leq 40$), 10.5 (if $S_{gn} \leq 200$), 21 (if $S_{gn} \leq 800$), 27 (otherwise)} \\
			$x_{d}''$ (p.u.) & [0.12; 0.3] & [0.09; 0.22] \\
			$x_{q}''$ (p.u.) & [$1 \times x_{d}''$; $3 \times x_{d}''$] & [$1 \times x_{d}''$; $1.1 \times x_{d}''$] \\
			$x_{d}'$ (p.u.) & [0.20; 0.45] (min $1.01 \times x_{d}''$) & [0.14; 0.35] (min $1.01 \times x_{d}''$) \\
			$x_{q}'$ (p.u.) & - & [0.9; 1.0] $\times x_{d}'$ (min $1.01 \times x_{q}''$) \\
			$x_d$ (p.u.) & [0.8; 1.4] (min $1.01 \times x_{d}'$) & [1.4; 3.0] (min $1.01 \times x_{d}'$) \\
			$x_q$ (p.u.) & [0.6; 0.7] $\times x_d$ (min $1.01 \times x_{q}''$) & [0.9; 1.0] $\times x_d$ (min $1.01 \times x_{q}'$) \\
			$T'_{d}$ (s) & [0.7; 2.2] & [0.8; 1.5] \\
			$T'_{q}$ (s) & - & [0.298; 0.756] \\
			$T''_{d}$ (s) & [0.02; 0.1] & [0.02; 0.05] \\
			$T''_{q}$ (s) & [0.017; 0.035] & [0.01; 0.035] \\
			$x_{(0)}$ (p.u.) & [0.05; 0.2] & [0.03; 0.1] \\
			$r_{(0)}$ (p.u.) & 0 & 0 \\
			$x_{(2)}$ (p.u.) & [0.9; 1.1] $\times x_{d}''$ & [0.9; 1.1] $\times x_{d}''$ \\
			$r_{(2)}$ (p.u.) & [0.014; 0.060] & [0.005; 0.029] \\
			$T_{ag}$ (s) & [5; 10] & [5; 10] \\
			$i_{sat}$ & 3 & 3 \\
			$S_{G1.0}$ (p.u.) & [0.064; 0.313] & [0.07; 0.340] \\
			$S_{G1.2}$ (p.u.) & [ 0.282; 1.018] (min $S_{G1.0}$) & [0.290; 1.12] (min $S_{G1.0}$) \\
			$r_{str}$ (p.u.) & [0.0014; 0.006] & [0.0016; 0.005] \\
			$x_l$ (p.u.) & [0.12; $\min(0.9, x_{d}'', x_{q}'')$] & [0.07; $\min(0.333, x_{d}'', x_{q}'')$] \\
			$q_{max}$ (p.u.) & \multicolumn{2}{c}{0.48\tnote{a}}  \\
			$q_{min}$ (p.u.) & \multicolumn{2}{c}{-0.41\tnote{b}} \\
			\bottomrule
		\end{tabular}
		\begin{tablenotes}
			\item[a] Corresponding to inductive $\cos\phi$ of 0.90\cite{schulzElektrischeEnergieversorgung2016}
			\item[b] Corresponding to capacitive $\cos\phi$ of 0.925\cite{schulzElektrischeEnergieversorgung2016}
		\end{tablenotes}
	\end{threeparttable}
	\label{tab:syn_mach}
\end{table*}

\begin{table*}[!h]
	\centering
	\caption{AC1A exciter configuration parameters (nominal values with $\pm$10\% variation unless specified otherwise)}
	\begin{threeparttable}
		\begin{tabular}{lc}
			\toprule
			\textbf{Parameter} & \textbf{Nominal Value} \\
			\midrule
			$K_a$ & 400 \\
			$K_c$ & 0.2 \\
			$T_a$ & 0.02 \\
			$K_f$ & 0.03 \\
			$T_f$ & 1 \\
			$T_e$ & 0.8 \\
			$K_e$ & 1 \\
			$K_d$ & 0.38 \\
			$E_1$ & 3.14 \\
			$S_{E1}$ & 0.03 \\
			$E_2$ & 4.18 \\
			$S_{E2}$ & 0.1 \\
			$V_{amin}$ & -14.5 \\
			$V_{amax}$ & 14.5 \\
			\midrule
			$T_b$ & 0\tnote{a} \\
			$T_c$ & 0\tnote{a}  \\
			$T_r$ & 0\tnote{a}  \\
			$V_{rmin}$ & -999\tnote{a}  \\
			$V_{rmax}$ & 999\tnote{a}  \\
			\bottomrule
		\end{tabular}
		\begin{tablenotes}
			\item[a] No Variation
		\end{tablenotes}
	\end{threeparttable}
	\label{tab:exciter_ac1a}
\end{table*}

\begin{table*}[!h]
	\centering
	\caption{IBR parameters by voltage level.}
	\label{tab:ibr}
	\begin{tabular}{llcc}
		\toprule
		\textbf{Parameter} & \textbf{Description} & \textbf{20 kV} & \textbf{110 kV} \\
		\midrule
		$S_{\mathrm{n}}$ (\si{\mega\volt\ampere}) & Rated apparent power IBR & [1; 15] & [20; 50] \\
		$P_{n}$ (\si{\pu}) & Active power setpoint IBR & [0.1; 0.9] & \\
		$Q_{n}/P_{n}$ (\si{\pu}) & Reactive power setpoint IBR & \multicolumn{2}{c}{[0.05; 0.1]} \\
		
		$S_{n,t}$ (\si{\mega\volt\ampere}) & Rated apparent power transformer & \multicolumn{2}{c}{Identical to IBR} \\
		$u_{\mathrm{k}}$ (\si{\percent}) & Short-circuit voltage transformer & 6 & 12 \\
		$P_{\mathrm{k}}$ (\si{\kilo\watt}) & Winding losses transformer & 18 & 130 \\
		\bottomrule
	\end{tabular}
\end{table*}

\begin{table*}[!h]
	\centering
	\caption{Grounding system configuration parameters by voltage level.}
	\begin{tabular}{lccc}
		\toprule
		\textbf{Parameter} & \textbf{20 kV} & \textbf{110 kV} & \textbf{345 kV} \\
		\midrule
		Grounding Type & \{solid, resistive, resonant\} & \{solid, resistive, resonant\} & solid \\
		ASC Tuning $I_{\mathrm{ASC}}$ (\si{\ampere}) & [-10; 10] & [-10; 10] & - \\
		Resistive Share of Coil Current (\%) & [2; 3] & [2; 3] & - \\
		Solid Grounding Resistance $R_{\mathrm{gs}}$ (\si{\ohm}) & [0.1; 2] & [0.1; 2] & [0.1; 2] \\
		Resistive Grounding Resistance $R_{\mathrm{gr}}$ (\si{\ohm}) & [5.77; 9.62] & [6.35; 31.75] & - \\
		\bottomrule
	\end{tabular}
	\label{tab:grounding}
\end{table*}

\begin{table*}[!h]
	\centering
	\caption{Shunt capacitor parameters by voltage level.}
	\begin{tabular}{lcc}
		\toprule
		\textbf{Parameter} & \textbf{20 kV} & \textbf{110/345 kV} \\
		\midrule
		Capacitor Power $Q_{\mathrm{c}}$ (\si{\mega\VAr}) & [0.3; 2.4] & [60.0; 200] \\
		\bottomrule
	\end{tabular}
	\label{tab:capacitors}
\end{table*}

\begin{table*}[!h]
	\centering
	\caption{Arc model parameters for cable faults and overhead line faults of different voltage levels.}
	\begin{tabular}{lcccc}
		\toprule
		\textbf{Parameter} & \textbf{Cable} & \textbf{20 kV} & \textbf{110 kV} & \textbf{345 kV} \\
		\midrule
		Arc time constant $\tau$ (\si{\milli\second}) & [0.05; 0.4] & [0.05; 0.65] & [0.3; 1.2] & [0.3; 1.2] \\
		Arc voltage $u_0$ (\si{\volt}) & [300; 4000] & [535; 1075] & [2200; 5800] & [4900; 13900] \\
		Arc resistance $r_0$ (\si{\ohm}) & [0; 0.015] & [0.014; 0.038] & [0.088; 0.248] & [0.208; 0.608] \\
		\bottomrule
	\end{tabular}
	\label{tab:arc_parameters}
\end{table*}

\FloatBarrier
\section*{Data Records}
The EvEMTBench dataset is compressed as \emph{.tar.gz} files. 11 such files can be downloaded directly from the data repository:
\begin{itemize}
	\item \textbf{File} \emph{adapt\_grid-CigreMVGrid.tar.gz}: Adaptgrid dataset containing randomly sampled events for the 20 kV CIGRE medium voltage benchmark grid with 14 buses.
	\item \textbf{File} \emph{adapt\_grid-DoubleLine.tar.gz}: Adaptgrid dataset containing randomly sampled events for the 110 kV test grid with parallel transmission lines.
	\item \textbf{File} \emph{adapt\_grid-TestGrid110kV.tar.gz}: Adaptgrid dataset containing randomly sampled events for the 110 kV test grid with multiple buses and lines.
	\item \textbf{File} \emph{adapt\_grid-IEEE39BusSystem.tar.gz}: Adaptgrid dataset containing randomly sampled events for the 345 kV IEEE 39-bus New England test system.
	\item \textbf{File} \emph{benchmark-CigreMVGrid.tar.gz}: Benchmark dataset containing discrete events for the 20 kV CIGRE medium voltage benchmark grid with 14 buses.
	\item \textbf{File} \emph{benchmark-DoubleLine.tar.gz}: Benchmark dataset containing discrete events for the 110 kV test grid with parallel transmission lines.
	\item \textbf{File} \emph{benchmark-TestGrid110kV.tar.gz}: Benchmark dataset containing discrete events for the 110 kV test grid with multiple buses and lines.
	\item \textbf{File} \emph{benchmark-IEEE39BusSystem.tar.gz}: Benchmark dataset containing discrete events for the 345 kV IEEE 39-bus New England test system.
	\item \textbf{File} \emph{multigrid-Template20kV.tar.gz}: Multigrid dataset containing events in synthetically generated 20 kV grid topologies.
	\item \textbf{File} \emph{multigrid-Template110kV.tar.gz}: Multigrid dataset containing events in synthetically generated 110 kV grid topologies.
	\item \textbf{File} \emph{multigrid-Template345kV.tar.gz}: Multigrid dataset containing events in synthetically generated 345 kV grid topologies.
\end{itemize}
\smallskip
After unpacking, each folder contains three subfolders: \emph{data}, \emph{labels}, and \emph{graphs}, organized as follows:
\begin{itemize}
	\item \textbf{Folder} \emph{data}: contains the simulation result files in CSV format.
	\begin{itemize}
		\item \textbf{File} \emph{result\#.csv}: for example, \emph{result0.csv} is a CSV file containing sub-millisecond-resolution three-phase voltage and current measurements from PowerFactory simulations. The CSV file contains a multi-index where the first level defines the measurement location and the second level defines the measured value.
		\begin{itemize}
			\item \textbf{Field} \emph{b:tnow in s}: time in seconds
			\item \textbf{Field} \emph{Measurement Location/Measured Value}: Phase voltages and phase currents at dedicated measurement locations. E.g. \textit{Cubicle(2)\textbackslash pex\_MainBus0\_MainLn\_0\_1\_0/c:Isec:A in A} contains the phase current A at MainBus0 of the transmission line MainLn\_0\_1\_0, which connects MainBus0 and MainBus1. All current measurements are oriented towards the branch element.
		\end{itemize}
	\end{itemize}
	\item \textbf{Folder} \emph{labels}: contains the metadata and simulation settings.
	\begin{itemize}
		\item \textbf{File} \emph{settings\_clean.csv}: the CSV file contains the configuration parameters for each simulation. It must be noted that all fields are determined for each simulation but may be ignored if they do not apply to the selected events (e.g. arc parameters are ignored for faults without arcs).
		\begin{itemize}
			\item \textbf{Field} \emph{general/result\_file\_path}: path to the corresponding result file.
			\item \textbf{Field} \emph{general/sim\_idx}: simulation index.
			\item \textbf{Field} \emph{general/vn}: nominal voltage level in kV (20, 110, or 345).
			\item \textbf{Field} \emph{general/fn}: nominal frequency in Hz (50 or 60).
			\item \textbf{Field} \emph{general/grid\_graph}: filename of the corresponding grid graph representation.
			\item \textbf{Field} \emph{events/event\_type}: type of event simulated (e.g., \emph{flt\_1phg\_shc} for a single line to ground fault).
			\item \textbf{Field} \emph{events/event\_start}: event start time in seconds.
			\item \textbf{Field} \emph{events/event\_target}: target component where the event occurs (e.g., \emph{MainLn1-2} for line between Bus1 and Bus2).
			\item \textbf{Field} \emph{events/event\_flt\_target\_line\_location}: fault location along the line in percentage (1.0 to 99.0).
			\item \textbf{Field} \emph{events/event\_flt\_shc\_resistance}: short-circuit resistance in Ohm.
			\item \textbf{Field} \emph{events/event\_flt\_hif\_resistance}: high-impedance fault resistance in Ohm.
			\item \textbf{Field} \emph{events/event\_phase\_select\_2ph}: phase selection for two-phase events (e.g., \emph{ab}).
			\item \textbf{Field} \emph{events/event\_phase\_select\_1ph}: phase selection for single-phase events (e.g., \emph{a}).
			\item \textbf{Field} \emph{events/event\_iflt\_duration}: fault duration in seconds.
			\item \textbf{Field} \emph{element/parameter}: additional parameters of different grid elements that are randomly sampled. For example arc/arc\_tau contains the time constant of the arc model.
		\end{itemize}
	\end{itemize}
	\item \textbf{Folder} \emph{graphs}: contains grid graph representations for machine learning applications.
	\begin{itemize}
		\item \textbf{File} \emph{graph\_grid\#.pickle}: pickled NetworkX graph object representing the grid topology with node and edge features.
		\item \textbf{File} \emph{grid\#.pfd}: PowerFactory object that can be directly imported into PowerFactory containing the grid topology.
	\end{itemize}
\end{itemize}

\section*{Technical Validation}
\begin{figure}
	\centering
	\includegraphics[width=0.9\linewidth]{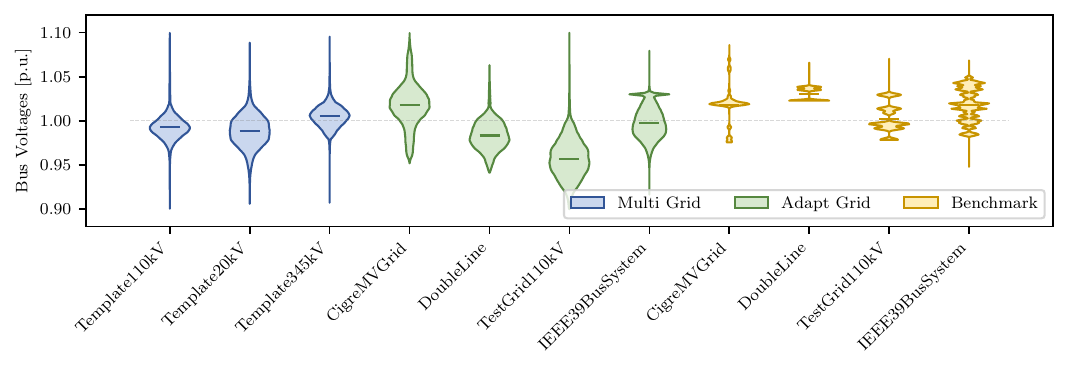}
	\caption{Distribution of RMS bus voltages during the first two cycles of the simulation for different dataset types and grids.}
	\label{fig:voltagermsviolin}
\end{figure}
\begin{figure}
	\centering
	\includegraphics[width=0.9\linewidth]{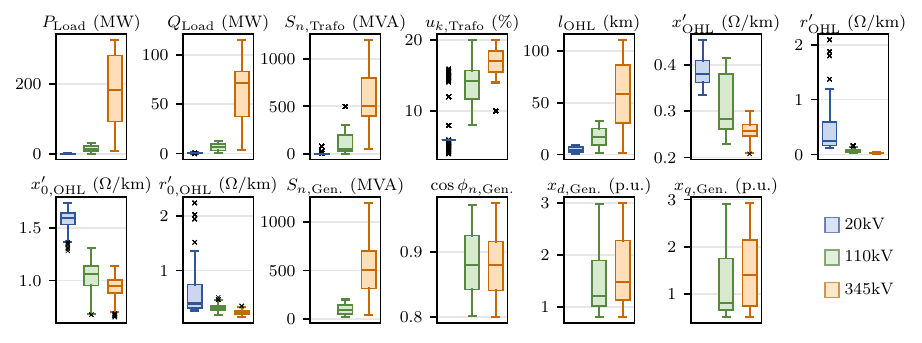}
	\caption{Selected distributions of parameters for Loads, Transformers, Overhead Lines and Synchronous Machines.}
	\label{fig:parameterranges}
\end{figure}
\begin{figure}
	\centering	\includegraphics[height=0.9\textheight]{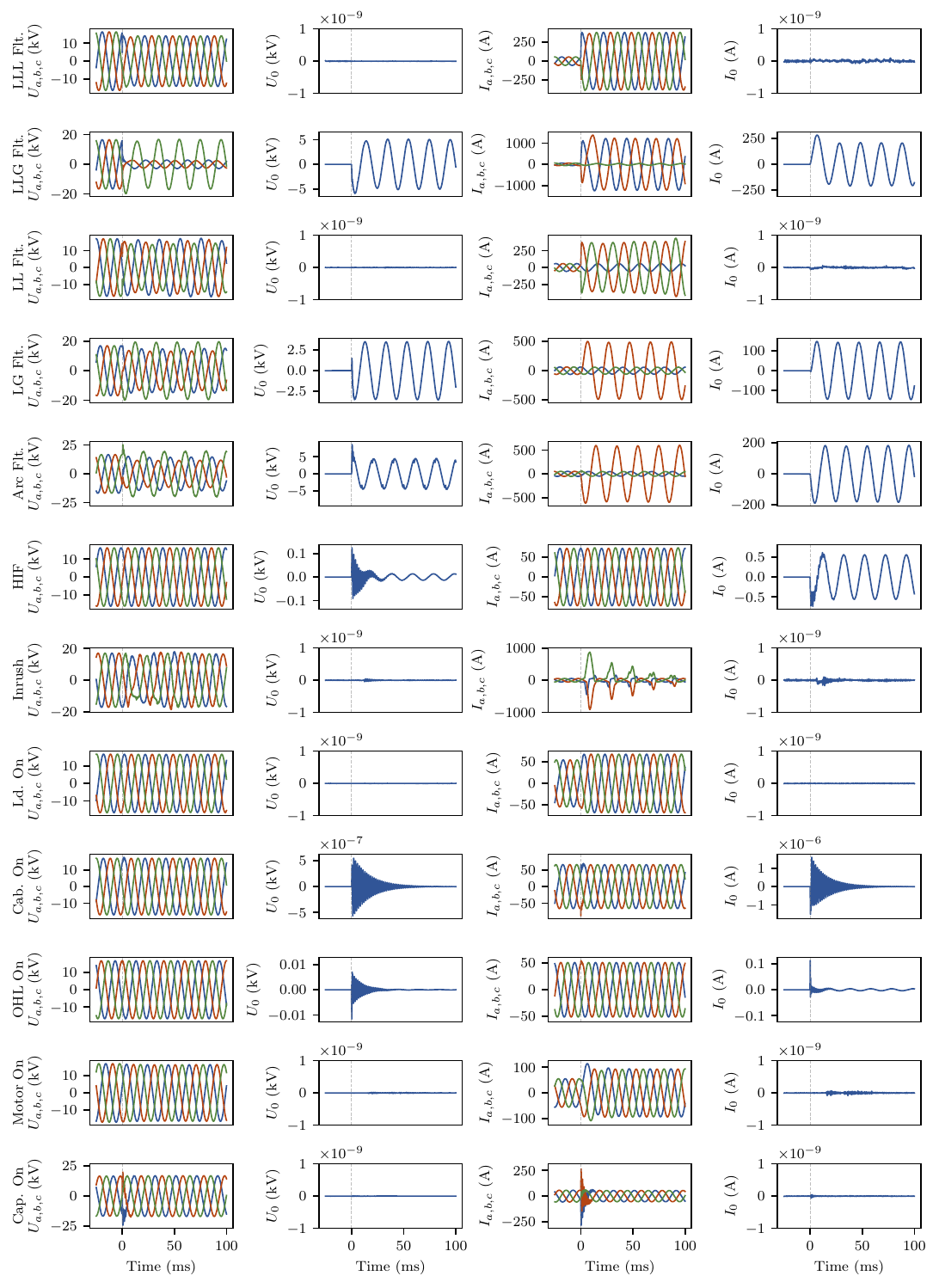}
	\caption{A selection of 12 simulated exemplary events recorded in the CIGRE MV Grid.}
	\label{fig:exampleevents}
\end{figure}
As the EvEMTBench dataset is simulation-based, trust in the technical validity of the data mainly arises from the detailed description of the dataset creation procedure in the Section \nameref{sec:methods}. Additionally, we provide visualizations of the dataset and the simulation parameters in \autoref{fig:voltagermsviolin}, \autoref{fig:parameterranges}, and \autoref{fig:exampleevents}. The goal of these visualizations is to illustrate the diversity and characteristics of the dataset to foster trust in the validity of the data.

In \autoref{fig:voltagermsviolin}, we analyze the root mean square (RMS) voltages of the first two cycles in each simulation. The visualization shows that a large majority of voltages is close to the nominal voltage of \qty{1.0}{\pu}. A smaller fraction of bus voltages exhibits greater variation, representing less common operating conditions of the grid. The TestGrid110kV represents a high load scenario, characterized by lower median voltage. The voltage distributions in the benchmark dataset are less continuous than the voltage distributions of the multigrid and adaptgrid datasets. This can be attributed to the fact that the simulation parameters in the benchmark dataset are not randomly sampled, but instead determined for discrete predefined values. Overall, the distribution of RMS voltages is consistent with realistic grid operating states.

Additionally, we visualize the distributions of important parameters of different grid elements in \autoref{fig:parameterranges}. Notably, the overhead line parameters are not set directly in the simulations, but are determined indirectly from the line geometry. The consistency with literature values confirms our approach. The high number of outliers for MV transformers is expected, as a large majority of transformers in \qty{20}{\kilo\volt} grids connects loads at lower voltage levels. Therefore, transformers connecting the \qty{20}{\kilo\volt} grid to higher voltage levels appear as outliers due to their smaller number and significantly higher short circuit voltage. Overall, the distributions of all parameters are consistent with previously cited literature. 

Finally, we present a small sample of fault and operating events in \autoref{fig:exampleevents}. For symmetric faults like LL and LLL faults, no zero sequence voltage or current can be observed. In contrast, LG and LLG faults induce significant voltages and currents in the zero sequence system. Cable and overhead line switches cause smaller transients. Additionally, high impedance faults are barely visible in the phase currents and voltages, but clearly visible in the zero sequence system. Load and motor switches cause increasing phase currents with different transients. Both transformer and capacitor switching events are clearly visible in the phase currents. Overall, the transients induced by these events match expected waveforms from literature.

\section*{Usage Notes}
Each individual dataset is compressed due to its size. We recommend extracting and starting with the simple double line grid model for initial experiments due to its simplicity and usability. To avoid data leakage between training and validation sets, we highly recommend simulation-level splits for the adaptgrid dataset and topology-level splits for the multigrid dataset. 

The dataset can be used for a wide range of power system protection related tasks, including fault detection, fault classification, fault line indentification, fault localization and fault area classification as defined by Oelhaf et al.\cite{oelhafScopingReviewMachine2025}. Additionally, the dataset can be used for broader tasks like event classification and localization as defined by Wilson et al\cite{wilsonGridEventSignature2024} or incipient fault detection as defined by Li et al.\cite{liIncipientFaultDetection2023}.

\section*{Data Availability}
The EvEMTBench dataset is publicly available and can be accessed via FAUDataCloud\cite{zotero-item-36862} using the following link: \url{https://data.fau.de/share/0e8d60feb7e65616c60aab78b93db77053275da53fd894bf5b75fc5e9ee7dfbf/}.

\newpage
\bibliography{MyLiterature}

\section*{Author Contributions}
\textbf{Georg Kordowich}: Writing -- original draft, Conceptualization, Methodology, Data Curation, Investigation. \\
\textbf{Jonathan Loebel}: Writing -- original draft, Methodology. \\
\textbf{Julian Oelhaf}: Writing -- original draft, Methodology. \\
\textbf{Andreas Maier}: Supervision, Project administration, Funding acquisition. \\
\textbf{Christian Bergler}: Writing -- review \& editing, Supervision. \\
\textbf{Siming Bayer}: Writing -- review \& editing, Supervision.\\
\textbf{Johann Jaeger}: Writing -- review \& editing, Methodology, Supervision, Project administration, Funding acquisition.

\section*{Competing Interests}
The authors declare no competing interests.

\section*{Funding}
This project was funded by the Deutsche Forschungsgemeinschaft (DFG, German Research Foundation) - 535389056.

\end{document}